\documentclass[twocolumn,apl,amsmath,amssymb,showpacs,superscriptaddress]{revtex4-2}
\usepackage{epsf} 
\usepackage{graphicx}
\usepackage{color}
\usepackage{soul}
\usepackage{gensymb}
\usepackage{sidecap}
\usepackage{amsmath}
\usepackage{mathtools}
\usepackage{float}
 \usepackage{multirow}
\usepackage[hidelinks,colorlinks=true,linkcolor=blue,citecolor=blue]{hyperref}
\DeclareUnicodeCharacter{2212}{\textendash}

\begin{document}

\title{Observation of an unconventional giant negative exchange bias effect in La$_{0.5}$Sr$_{0.5}$Co$_{0.85}$Nb$_{0.15}$O$_3$} 

\author{Rishabh Shukla}
\affiliation{Department of Physics, Indian Institute of Technology Delhi, Hauz Khas, New Delhi-110016, India}
\author{B. Schwarz}
\affiliation{Institute for Applied Materials (IAM), Karlsruhe Institute of Technology (KIT), 76344, Eggenstein-Leopoldshafen, Germany}
\author{R. S. Dhaka}
\email{rsdhaka@physics.iitd.ac.in}
\affiliation{Department of Physics, Indian Institute of Technology Delhi, Hauz Khas, New Delhi-110016, India}

\date{\today}

\begin{abstract}

We find an unconventional giant negative exchange bias (EB) of $H_{\rm EB}$ = --14.1~kOe at 2~K (cooling field of 50~kOe) in the cluster spin-glass (CSG) La$_{0.5}$Sr$_{0.5}$Co$_{0.85}$Nb$_{0.15}$O$_3$ perovsikte cobaltites. The magnetic memory effect, aging measurements, and nonlineraity in specific heat capacity reveal the glassy magnetic state at low temperatures. Further, the detailed analysis of {\it ac-}magnetic susceptibility confirms the glassy state below $\sim$58~K and the obtained characteristic spin-relaxation time-scale of $\tau_0$ = 8.4$\times$10$^{-10}$~s indicates the presence of CSG. Moreover, the analysis of magnetic training effect using the classical EB relaxation model reveals that the frozen spins relax slowly as compared to the rotatable spins at the interface of antiferromagnetic/ferromagnetic (AFM/FM) regions in CSG. Interestingly, the dependence of EB parameters is found to be unconventional for cooling field $>$50~kOe as the $H\rm_{EB}$ and $M\rm_{EB}$ show decreasing trend instead of expected saturation at higher fields. This unusual nature emerges due to large negative values of intrinsic interface exchange coupling ($J_i$), i.e., --10.24$\pm$0.22~meV and --12.55$\pm$0.49~meV for the measuring fields of $\pm$50~kOe and $\pm$90~kOe, respectively, whereas the number of spins in the FM cluster ($N_{\rm FM}$) are found to be small in the range of 2.4--3.1. These obtained values of $J_i $ and $N_{\rm FM}$ indicate the dominant AFM interactions and the presence of FM clusters in the AFM matrix, respectively, which correlate well with the observed unconventional behavior of giant negative exchange bias in the present sample. 

\end{abstract}

\maketitle

\section{\noindent ~Introduction}

The essence of futuristic magnetic materials having large values of saturation magnetization and coercive fields have been demonstrated through the artificially developed exchange bias (EB) mechanism \cite{MeiklejohnPR56+57} for the magnetic storage devices, spin-valves, high-density magnetic recordings, domain stabilizers, permanent magnets, etc. \cite{DienyJMMM94, RaduPRB09, TsangIEEE82, DienyPRB91}. The EB phenomenon was first discovered in the magnetic nanostructures of Co nanoparticles coated with a layer of CoO \cite{MeiklejohnPR56+57}. Further, the EB effect is observed in the thin-film heterostructures via coupling of ferromagnetic (FM) and antiferromagnetic (AFM) layers, inhomogeneous spin-glasses via coupling of FM and AFM domains, artificial FM/AFM superlattices, etc. \cite{SkumryevNature03, VelthuisAPL99, NoguesPRL96, AliNM07}. In general, the EB effect is understood in terms of the interface exchange coupling between FM and AFM layers, which leads to the unidirectional exchange anisotropy, and a shift in the isothermal magnetization (M--H) loops is observed on cooling the sample below the N\'eel temperature in the presence of an applied magnetic field \cite{MeiklejohnPR56+57, NoguesPRL96, StampsJPD00, PengPRB00}. Moreover, this understanding was further extended to the EB effect observed at the interface of FM--spin glass (SG), FM--ferrimagnet (FIM), AFM--SG, AFM--FIM--SG, as well as solely in cluster spin glass (CSG) system \cite{ManivNP21, DingPRB13, KarmakarPRB08}. However, the magnitude and characteristics of the observed EB effect are greatly influenced by the choice of interface, type of magnetic interactions, thickness of the interface, etc. \cite{RaduPRB09, StampsJPD00}. For example, in the AFM--FM interface (since FM and AFM have small and large anisotropy, respectively), on cooling the sample below T$\rm_N$ the spins orient towards the applied magnetic field; however, at the interface, the spins exhibit a unidirectional anisotropy and a large enough measuring field overcomes the interface magnetization and reverses the FM \cite{RaduPRB09}. These can result in shifting the hysteresis loop by the magnitude of the interlayer exchange coupling opposite to the applied magnetic field \cite{AntelPRL99, OhldagPRL03}. This suggests that an interface is required to induce the pinning effect for the observation of the EB. This pinning effect need not originate because of the uncompensated spins, but it can also emerge due to the presence of frozen spins, very likely in the glassy phases \cite{GruytersPRB00}. 

Note that, for effective utilization of the EB effect, a comprehensive understanding of the EB effect in terms of interface magnetism, disorder, inhomogeneous magnetization, etc. is vital and plays a crucial role in designing the materials of desirable applications \cite{Ajay_PRB24, NoguesPRL96, StampsJPD00, GiriJPCM11}. Moreover, the use of SG as an essential ingredient in these artificial EB systems is very useful to widen the understanding of the effects of frozen spins, frustration, and inhomogeneous magnetic regions. In this line, the SG-supported EB systems are also technologically useful due to their presence in magnetically disordered materials and their composites \cite{WangPRB04, ZhangAPL04}. In order to explain the SG phenomenon, the droplet model \cite{FisherPRB88, AnderssonPRB93} and hierarchical models \cite{LeflochEL92} are used, which utilize the multi-valley energy structure and equivalent spin configurations. Despite several efforts to understand the EB effect in single-phase magnetically inhomogeneous compounds \cite{ManivNP21, LiaoAPR23} the understanding of the exchange bias phenomenon in SG phases is elusive. Therefore, a complete picture of the spin-glass driven EB is vital for engineering the spintronics devices, magnetic memory devices, and other technological advances \cite{AliNM07, MeirzadehACSCS2021}. In transition metal oxide the EB has been investigated under different schemes, which involve, thin film interfaces, core-shell nanoparticles, cationic substitution in magnetically ordered systems, etc. \cite{SunNL12, SanthoshPRB18, CourtimPRB16, PrajapatJMMM19}. However, the observation of EB and its origin in polycrystalline samples remain debatable. For example, in cobaltites, the observed magnitude of EB varies significantly, i.e., 5.5~kOe in SrLaCo$_{0.5}$Mn$_{0.5}$O$_4$ \cite{SanthoshPRB18}, 1.4~kOe in Sr$_{1.5}$Pr$_{0.5}$CoO$_4$ \cite{AngJAP08}, 1.1~kOe in La$_{1.5}$Ca$_{0.5}$CoIrO$_6$ \cite{CourtimPRB16}, 1.2~kOe in LaSrCoFeO$_6$ \cite{SahooPRB19}, 2.1~kOe in La$_{1.5}$Ca$_{0.5}$CoFeO$_6$ \cite{SinghJPCM22}, 5.1~kOe in Sm$_{1.5}$Ca$_{0.5}$CoMnO$_{6}$ \cite{GiriJPD16}, 6~kOe and 9.5~kOe in SrLaFe$_{0.25+x}$Mn$_{0.25}$Co$_{0.5-x}$O$_4$ \cite{AnusreePRB20}, 8.22~kOe in SrCo$_{1-x}$V$_x$O$_{3-\delta}$ \cite{PrajapatJMMM19}, 0.8~kOe in Sr$_2$CoNbO$_6$ \cite{Ajay_PRB24}, etc.

Furthermore, the hole doping of divalent cations in LaCoO$_3$, i.e., La$_{1-x}$Sr$_x$CoO$_3$ enthralled the researchers due to the observed colossal magnetoresistance (CMR), which is believed to have originated from the presence of magnetic nonhomogeneity in the samples \cite{GolovanovPRB96, TangPRB06, WuPRB05}. The substitution of diamagnetic Nb$^{5+}$ (4$d^0$) in La$_{0.5}$Sr$_{0.5}$Co$_{1-x}$Nb$_x$O$_3$ converts the Co$^{4+}$ ions into Co$^{3+}$ ions, and therefore an increase in the density of Co$^{3+}$ over Co$^{4+}$ ions \cite{ShuklaPRB23, ShuklaPRB18}. The FM clusters appear due to the double exchange (DE) interactions from the Co$^{4+}$-O-Co$^{3+}$, whereas the AFM clusters emerge due to the super-exchange (SE) interactions from Co$^{3+}$-O-Co$^{3+}$ and Co$^{4+}$-O-Co$^{4+}$ \cite{ZenerPR51_1, ZenerPR51_2}. The electrons in the outer shell of transition metals are localized, and hence the interactions strongly depend on the orientation of the localized spin moments \cite{ZenerPR51_1, AndersonPR55}. Also, the observed magneto-electronic intrinsic phase separation (MIPS) plays a crucial role in the generation of the EB in these complex oxides owing to the presence of hole-rich metallic FM clusters and hole-poor insulating AFM clusters \cite{KuhnsPRL03, HochPRB04}. 

Therefore, we investigate the evolution of exchange bias (EB), memory effect and heat capacity in bulk La$_{0.5}$Sr$_{0.5}$Co$_{0.85}$Nb$_{0.15}$O$_3$ sample. The {\it ac-}magnetic susceptibility data confirms the glassy nature of the sample below $\sim$58~K (T$\rm_f$) and the estimated characteristic time-scale of $\tau_0$ = 8.4$\times$10$^{-10}$~s through detailed analysis suggests that the individual spin entities behave like the cluster spin-glass (CSG). Moreover, the nonlinearity in specific heat capacity at low-temperature is analyzed including the magnetic contributions and further validates the glassy magnetic state. Intriguingly, we find an unconventional behavior of giant negative exchange bias -14.1~kOe (at 2~K) and -7.4~kOe (at 5~K) for a 50~kOe cooling field. The exchange interactions in CSG at the interface of FM and AFM clusters originated due to the hole-rich and hole-poor regions, respectively. The exchange bias parameters (H$\rm_{EB}$ and M$\rm_{EB}$) decrease at the higher cooling fields ($>$50~kOe) which are modeled with the intrinsic interface exchange coupling and reveals $J_i =$ -10.24$\pm$0.22~meV and $N_{FM} =$ 2.4$\pm$0.2 (measuring field of $\pm$50~kOe) and $J_i =$ -12.55$\pm$0.49~meV and $N_{FM} =$ 3.1$\pm$0.1 (measuring field of $\pm$90~kOe), which is responsible for the reduction in the EB parameters at higher cooling fields. 

\section{\noindent ~Experimental Details}

Polycrystalline sample La$_{0.5}$Sr$_{0.5}$Co$_{0.85}$Nb$_{0.15}$O$_3$ was synthesized using the solid-state reaction route, where stoichiometric amount of as purchased SrCO$_3$, Co$_3$O$_4$, Nb$_2$O$_5$, and La$_2$O$_3$ (dried at 900$\rm^o$C for 6~hrs) were mixed homogeneously using an agate mortar-pestle and then heated in air using a muffle furnace (from Nabertherm) at 1000$\rm^o$C (48~hrs) and 1300$\rm^o$C (36~hrs) with intermittent grindings. The basic characterizations including structural, magnetization, transport and electronic properties of this sample are reported in \cite{ShuklaPRB23}. The detailed magnetic and specific heat measurements in different protocol are performed using the DynaCool Physical Property Measurement System from Quantum Design, USA at Karlsruhe Institute of Technology (KIT), Germany.

\section{\noindent ~Results and Discussion}

\begin{figure}[h]
\includegraphics[width=3.4in]{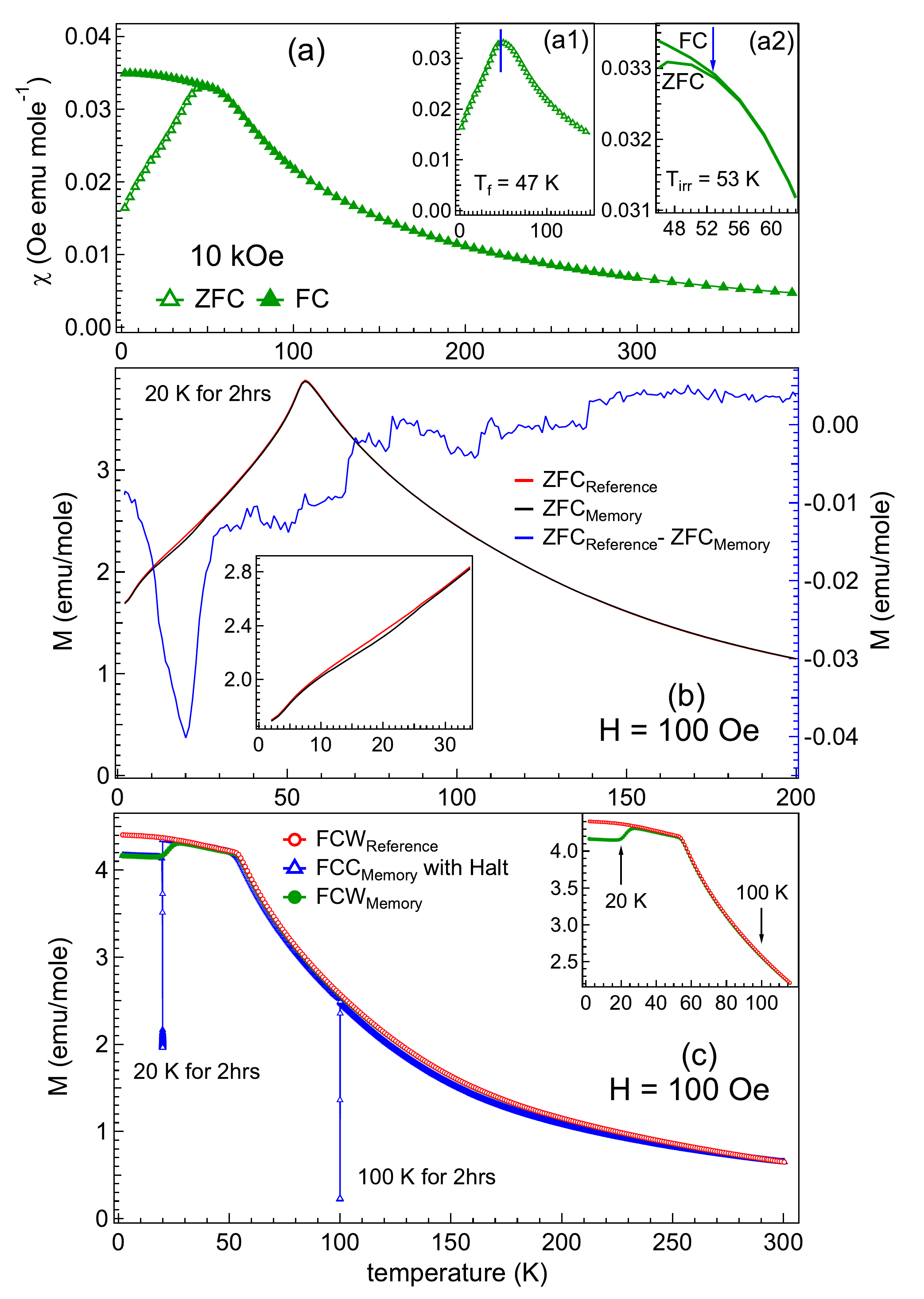}
\caption{(a) The zero-field cooled (ZFC) and field cooled (FC) magnetization curve recorded at 10~kOe. The insets show the zoomed view near (a1) T$\rm_f=$ 47~K and (a2) T$\rm_{irr}=$ 53~K. The magnetic memory effect in magnetization under (b) ZFC and (c) FC protocols at 100~Oe. Note that halts at 20~K and 100~K were employed during the measurement for 2~hrs each. The inset in (b) highlights  the memory effect under ZFC protocol, and inset in (c) highlights the magnetization drop in FCW at 20~K ($<$T$\rm_f$) and no change at 100~K ($>$T$\rm_f$). }
\label{fig:ZFC+FC+Memory}
\end{figure}

In Fig.~\ref{fig:ZFC+FC+Memory}(a) we show the temperature-dependent {\it dc-}magnetic susceptibility ($\chi\rm_{dc}$) of La$_{0.5}$Sr$_{0.5}$Co$_{0.85}$Nb$_{0.15}$O$_3$ measured at 10~kOe in both  zero-field-cooled (ZFC) and field-cooled (FC) protocols. The $\chi\rm_{dc}$ exhibits a monotonous increase with lowering the temperature till $\approx$47~K (termed as freezing temperature, T$\rm_f$); however, below T$\rm_f$, the $\chi\rm_{dc}$ decreases in the ZFC mode, while increases at a slower rate in the FC mode. A clear bifurcation/separation in the ZFC-FC curves near 53~K (termed as irreversibility temperature, T$\rm_{irr}$) and the observed cusp/peak in the ZFC curve are resultant of the competition between the energy of anisotropic field [due to the ferromagnetic (FM)/antiferromagnetic (AFM) interactions] and the applied field energy \cite{HoJMMM19}, which is considered to be a favorable condition for the spin glass (SG) behavior \cite{BagPRB18, Kumar_PRB1_20} and exchange bias \cite{ManivNP21}. Also, the observed shift in the T$\rm_f$ value from 55~K at 100~Oe \cite{ShuklaPRB23} to 47~K at 10~kOe (see inset) advocates for the presence of SG state below the T$\rm_f$ in the sample. To understand further, we perform the memory effect measurements under the ZFC and FC protocols where initially the sample is cooled from 300~K to 2~K with a cooling rate of 2~K/min in the absence of the magnetic field without any halts and then measured the magnetization with the conventional method (during warming) at an applied field of 100~Oe with a heating rate of 2~K/min up to 300~K [this ZFC curve is marked as reference curve (solid red line) in Fig.~\ref{fig:ZFC+FC+Memory}(b)]. Now again the sample was cooled from 300~K to 2~K in the absence of a magnetic field at 2~K/min rate, but with the intermediate halts at 100~K ($>$T$\rm_f$) and 20~K ($<$T$\rm_f$) for 2~hrs each, and then measured the magnetization during warming at 100~Oe [this ZFC curve is marked as memory curve (solid black line) in Fig.~\ref{fig:ZFC+FC+Memory}(b)]. A shallow dip is observed for the memory curve near the 20~K halt temperature (below T$\rm_f$) even in the absence of a magnetic field; however, no difference in the curves near the 100~K halt temperature (above T$\rm_f$). This dip in the ZFC memory curve with respect to the ZFC reference curve is highlighted in the inset of Fig.~\ref{fig:ZFC+FC+Memory}(b). 
Moreover, the difference curve between the reference and memory ZFC curves exhibit a dip near 20~K only, which confirms that the system keeps the sample's history and indicates the glassy nature below the T$\rm_f$. Further, we measure the memory effect under the FC protocol, where initially sample was cooled from 300~K to 2~K with a magnetic field of 100~Oe at 2~K min$^{-1}$ rate without any halts, and measured the magnetization during warming with a heating rate of 2~K min$^{-1}$ [this FCW curve is marked as reference curve (open red circles) in Fig.~\ref{fig:ZFC+FC+Memory}(c)]. Then, the sample was cooled from 300~K to 2~K with a magnetic field of 100~Oe at a rate of 2~K min$^{-1}$ including the intermediate halts at 20~K ($<$T$\rm_f$), and 100~K ($>$T$\rm_f$) for 2~hrs each. Note that, the magnetic field is switched off during the halt period and then switched on during the subsequent sample cooling. In this case, the magnetization is measured during sample cooling and warming, and these curves are termed as FCC (open blue triangles) and FCW (solid green circles) in Fig.~\ref{fig:ZFC+FC+Memory}(c), respectively. The memory effect prevails below T$\rm_f$ where the magnetization decreases near 20~K, while above T$\rm_f$ there is no difference around 100~K, see the inset for clarity. These above observations suggest the presence of competitive/coupled FM--AFM interactions in the sample resulting the glassy magnetic state.

Now, first we investigate the spin dynamics of the glassy magnetic state through detailed analysis of the ac-magnetic susceptibility ($\chi\rm_{ac}$) data in terms of the real ($\chi$') and imaginary ($\chi$'') parts, as shown in Figs.~S1(a, c) and Figs.~S1(b, d) of \cite{SI}, respectively. The peak value in $\chi$' is found to be 58.2~K at 193~Hz excitation frequency, which shifted to 60~K at 9984~Hz along with the decrease in the magnitude of the magnetic susceptibility. Here, we first plot the variation of peak temperature with the excitation frequency (not shown) and extract the Mydosh parameter using the formula \cite{MydoshTF93}: $\delta{T_f} = (\Delta{T_f})/[T_f\Delta (log_{10}f)]$ where T$_f$ is the freezing temperature and $f$ is the excitation frequency. The obtained value of $\delta{T_f}=$ 0.017 is found to be higher as compared to the canonical spin glasses (AuMn, $\delta{T_f}$ = $\sim$0.0045 \cite{MulderPRB82} and CuMn, $\delta{T_f}$ = $\sim$0.004 \cite{MydoshTF93}) and lower as compared to the superparamagnets ($\alpha$-Ho$_2$O$_3$(B$_2$O$_3$), $\delta{T_f}$ = $\sim$0.28 \cite{MydoshTF93}). However, we note that the value for the present sample lies within the range of the cluster spin glasses (0.01--0.018) \cite{ChakrabartyJPCM14, ShuklaJPCC19}. Also, the obtained values of characteristic spin relaxation time from the analysis of aging measurements found to be consistent with glassy systems \cite{PakhiraPRB16, KhanPRB14}, as shown in Fig.~S3 and Table S1 of \cite{SI}. 

\begin{figure}[h]
\includegraphics[width=3.3in]{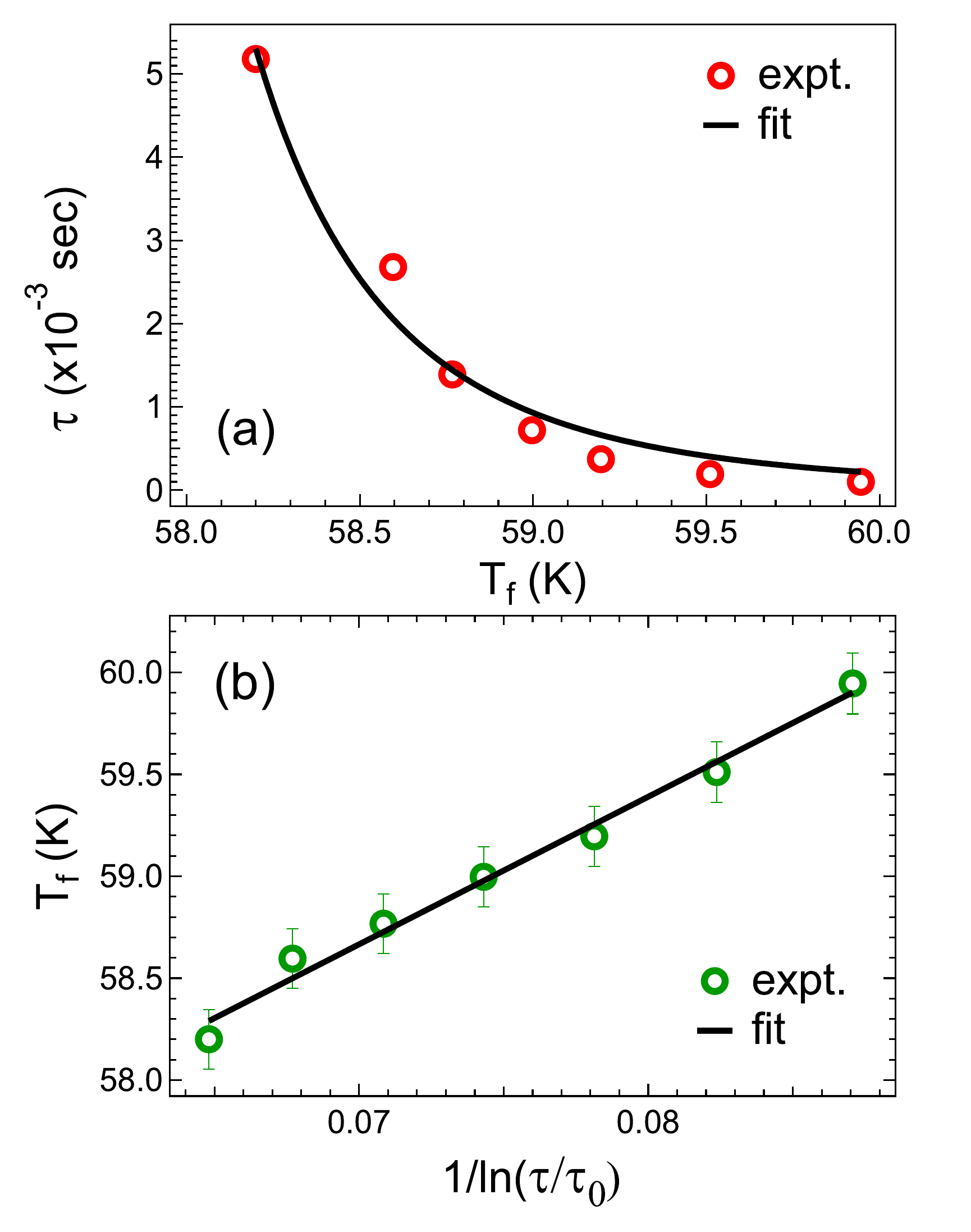}
\caption{(a) The dependence of the spin relaxation time ($\tau$) on the freezing temperature (T$\rm_f$) with an exponential fit (solid black line) using the dynamical scaling power law, and (b) the dependence of freezing temperature T$_f$ on 1/ln($\tau/\tau_0$) with a linear fit (solid black line) following the Vogel-Fulcher law. }
\label{fig:ac_A}
\end{figure}

Further, to extract the time scale of spin relaxation and type of interaction between magnetic clusters, the frequency dependence of the T$_f$ can be further analyzed with the dynamic scaling theory using the power law  \cite{SouletiePRB85},
\begin{eqnarray}
\tau = \tau_0 \left(\frac{T_F-T_{SG}}{T_{SG}}\right)^{-z\nu'}
\end{eqnarray}
where $\tau$ is the relaxation time corresponding to the measured frequency ($f=$ 1/$\tau$), $\tau_0$ is the characteristic relaxation time for the flipping of a single spin entity, T$\rm_{SG}$ is the spin-glass temperature or termed as the freezing temperature when $\tau$ diverges or frequency tends to zero, and z$\nu'$ is the dynamic critical exponent [where critical exponent, $\nu'$ is associated with the correlation length $\xi$ = (T$_f$/T$_{SG}$-1)$^{\nu'}$ and dynamical critical exponent, $z$ relates correlation length to spin relaxation time such that $\tau\sim\xi^z$]. The best fit (solid black line) to the experimental data using equation~1, as shown in Fig.~\ref{fig:ac_A}(a), yields the values of $\tau_0$ = 8.4$\pm$0.3$\times$10$^{-10}$~s, T$\rm_{SG}$ = 56.5$\pm$0.1~K, and z$\nu'$ = 4.5$\pm$0.4. For the canonical and cluster spin glasses, the $\tau_0$ exists in the range of 10$^{-12}$ to 10$^{-14}$~s and 10$^{-6}$~s to 10$^{-10}$~s, respectively, and z$\nu'$ lies between 4 and 12 \cite{ShuklaJPCC19, AnandPRB12}. In the present case, the obtained value of $\tau_0$ confirms the presence of cluster spin glass and indicates the slower spin dynamics. Also, we investigate the spin-interaction by the Arrhenius relation, as shown in Fig.~S2(a) of \cite{SI}, which manifests the presence of interaction between the individual spins (cluster spin glass). Moreover, the dynamical scaling behavior of spin-freezing is modified to the Vogel-Fulcher law, as given below  \cite{MydoshTF93, ShuklaJPCC19, SouletiePRB85},
\begin{eqnarray}
\tau = \tau_0 exp\left(\frac{-E_a}{k_B(T_f-T_0)}\right)
\end{eqnarray}
where T$_0$ is the characteristic temperature/Vogel-Fulcher temperature utilized to measure the strength of interaction between magnetic entities and other parameters are same as defined earlier. A simplified form of Vogel-Fulcher law is used to plot a graph between T$\rm_f$ versus 1/ln($\tau/\tau_0$) in Fig.~\ref{fig:ac_A}(b) and a best fit (solid black line) to the experimental data using equation~2 results in the values of E$_a$ = 6.8$\pm$1~meV, and T$_0$=53.0$\pm$0.5~K. Here, a non-zero value of T$_0$ (comparable to the freezing temperature, T$\rm_f$) in the present sample further validate the interaction between the magnetic clusters \cite{AnandPRB12}. 

\begin{figure*}
\includegraphics[width=7.0in, height = 4.3in]{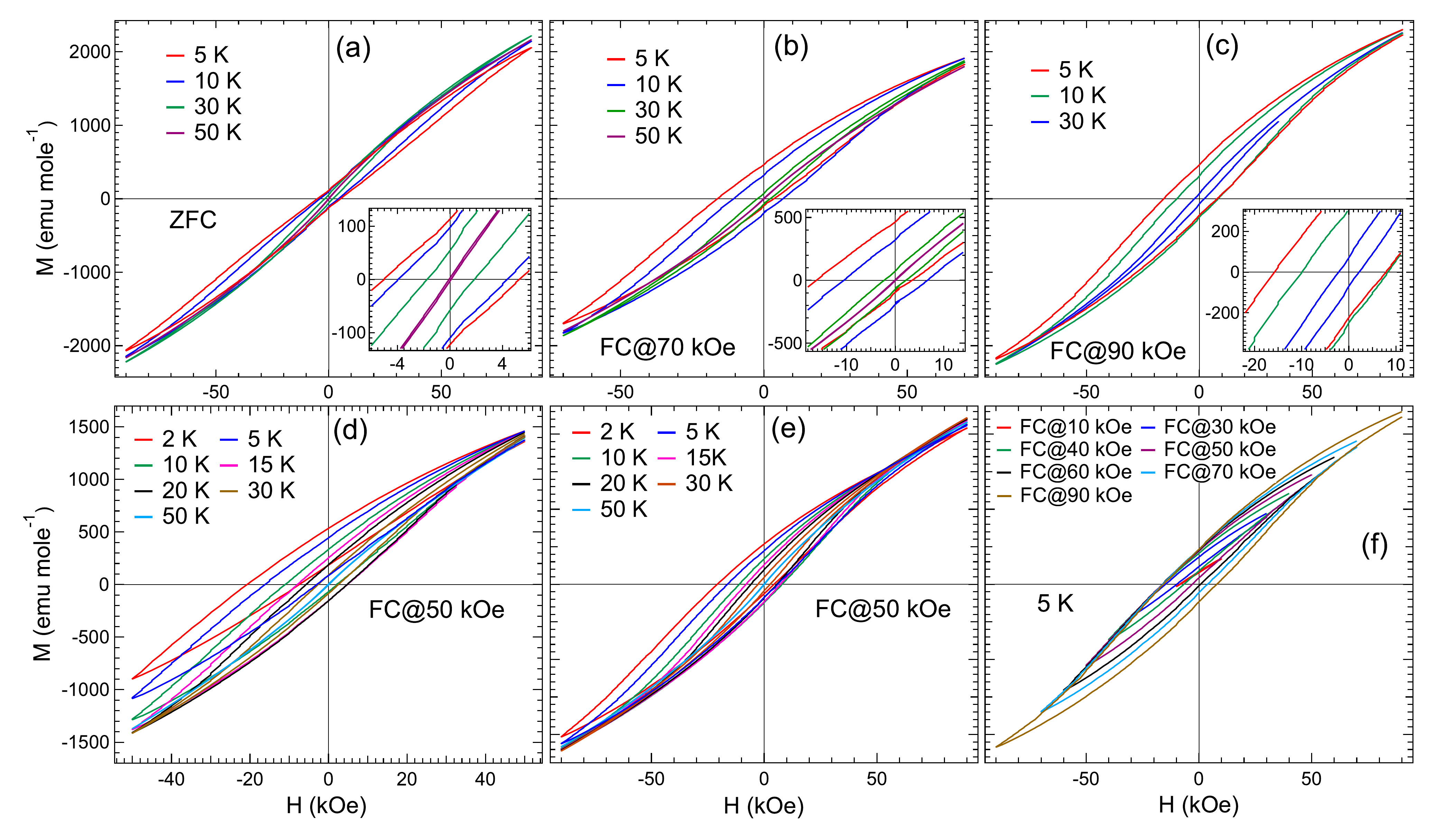}
\caption{The isothermal M--H loops recorded under the (a) ZFC protocol (b) FC protocol with a cooling field of 70~kOe and within $\pm$70~kOe, (c) with a cooling field of 90~kOe and within $\pm$90~kOe, (d) with a cooling field of 50~kOe and within $\pm$50~kOe, (e) with a cooling field of 50~kOe and within $\pm$90~kOe, and (f) a comparison of the M--H loops measured at different cooling fields and between different measuring fields at 5~K; insets in (a--c) highlight the zoomed view near origin.}
\label{fig:MH}
\end{figure*}

It is interesting to note here that the competitive AFM-FM interactions resulting in the cluster spin glass behavior can give rise to the exchange bias (EB) effect in the sample. Therefore, in order to get a complete picture of FM-AFM interactions and to investigate the EB effect we perform the isothermal field-dependent magnetization (M--H) below the T$\rm_f$ under different cooling protocols. In this line, first, we recorded the isothermal M--H loops under the ZFC protocol with the maximum applied magnetic field of $\pm$90~kOe at 5~K, 10~K, 30~K, and 50~K and a comparison of these curves is presented in Fig.~\ref{fig:MH}(a) where the inset showed the zoomed view within $\pm$6~kOe. Here, we define the coercivity $\rm H_{C} = \rm(|H_{C_1}|+|H_{C_2}|)/2$, where H$\rm_{C_1}$ and H$\rm_{C_2}$ are the left (-ve H-axis) and right (+ve H-axis) coercive fields; and retentivity $\rm M_{r} = \rm(|M_{r_1}|+|M_{r_2}|)/2$, where M$\rm_{r_1}$ and M$\rm_{r_2}$ are the top (+ve M-axis) and bottom (-ve M-axis) magnetic retentivity values. The estimated values of H$\rm_C$ and M$\rm_r$ are found to be 5.2~kOe and 129~emu mole$^{-1}$ at 5~K; 4.2~kOe and 104~emu mole$^{-1}$ at 10~K; 1.8~kOe and 56~emu mole$^{-1}$ at 30~K; and 84~Oe and 3~emu mole$^{-1}$ at 50~K, respectively. Interestingly, the values of H$\rm_C$ and M$\rm_r$ increase monotonously with a decrease in the temperature for T$<$T$\rm_f$, which indicate the dominance of FM interactions at low temperatures. Further, to elucidate the phenomenon of exchange bias we measure the isothermal M--H loops under the FC protocol in the cooling field of $\pm$70~kOe at 5~K, 10~K, 30~K, and 50~K, as shown in Figs.~S4(a--d) of \cite{SI}, respectively. These results for $\le$ 30~K manifests the shift in the M--H loop in opposite direction of the cooling field, but equal magnitude; however, no shift is observed at 50~K (see Fig.~S4 of \cite{SI}). This observed exchange bias effect below T$\rm_f$ affirms the intrinsic coupling between the FM--AFM interactions present in the sample. Another important observation is that there is no opening in the M--H loops in the high-field region when recorded under the ZFC protocol at different temperatures [see Fig.~\ref{fig:MH}(a)]; however, in case of the FC protocol we find an opening in the M--H loops at high-field region and have the correlation with the direction of the cooling field [see Fig.~S4 of \cite{SI}]. This opening in the M--H loops can be ascribed due to the time-dependent variation of the magnetization at high applied magnetic field. Further, we show the isothermal M--H loops under the FC protocol at various temperatures in Figs.~\ref{fig:MH}(b) for the cooling field of 70~kOe (measuring field of $\pm$70~kOe) and in Figs.~\ref{fig:MH}(c) for the cooling field of 90~kOe (measuring field of $\pm$90~kOe) where the respective insets compare the zoomed view near origin. Moreover, we can estimate the exchange bias field (H$\rm_{EB}$) and exchange bias coercivity (M$\rm_{EB}$) from the isothermal M--H loops recorded under the FC protocol using $\rm H_{EB} = \rm (H_{C_1}+H_{C_2})/2$ and $\rm M_{EB} = \rm (M_{r_1}+M_{r_2})/2$. We compared these parameters for different cooling fields at few temperatures, the H$\rm_C$ and M$\rm_r$ in Fig.~S5(a) of \cite{SI}, and the H$\rm_{EB}$ and M$\rm_{EB}$ in Fig.~S5(b) of \cite{SI}. We find that the H$\rm_C$ and M$\rm_r$ show nearly a linear decrease; while the H$\rm_{EB}$ and M$\rm_{EB}$ exhibit a logarithmic decrease as an increase in temperature affects the alignment of the spins significantly even below the T$\rm_f$ and almost randomizes the spins above the T$\rm_f$ (discussed in detail later). 

Moreover, the isothermal M--H loops are measured under the FC protocol with a constant cooling field of 50~kOe between 2--50~K, by varying the measuring fields within $\pm$50~kOe and $\pm$90~kOe, as shown in Figs.~\ref{fig:MH}(d, e), respectively. First we find a significant effect of measuring/cooling field; therefore, a detailed comparison of the M--H loops is shown in Figs.~S6(a--f) of \cite{SI} at various temperatures. Since the measuring field can be understood as a force to align the frozen/blocked spins along the field direction, and therefore, we observe a lower reduction of exchange bias parameters at higher temperatures as compared to low temperatures. In Fig.~S6(a) of \cite{SI}, we find that the M--H loop measured at 2~K shifted completely to the -ve H-axis and +ve M-axis, which manifests that $\pm$50~kOe magnetic field is not sufficient to realign the frozen spins, while a measuring field of $\pm$90~kOe seems to be enough to realign the frozen spins in the field direction (see Fig.~S6(a) of \cite{SI}). This indicates that the measuring field range also plays an important role in the behavior of the exchange bias effect. The values of H$\rm_{EB}$ (M$\rm_{EB}$) and H$\rm_{C}$ (M$\rm_{r}$) as highlighted in Figs.~S6(a) of \cite{SI} are estimated via the average value of the field where curves intercept the H-axis (M-axis) and half-width of the loop where curve intercepts the M-axis (H-axis), respectively. These extracted exchange bias parameters are summarized in Table S2 of \cite{SI}, where we observe very large values of H$\rm_{EB}$ = -14.1~kOe and M$\rm_{EB}$ = 359.6~emu mole$^{-1}$, which are found to be highest reported for the perovskite cobaltites in the literature as per the best of our knowledge. The values of H$\rm_{EB}$ and M$\rm_{EB}$ decreases to -8.23~kOe and 206.4~emu mole$^{-1}$, respectively, for the measuring field of $\pm$90~kOe. We also find decrease in the values of these parameters with increasing the temperature. Further, to understand the exchange bias behavior we measure the M--H loops at 5~K at different measuring and cooling fields in such a way that we first cool the sample in a constant magnetic field and then the M--H loops are measured within the $\pm$cooling field, as shown in Fig.~\ref{fig:MH}(f). In order to highlight this effect, we compare the isothermal M--H loops in Figs.~S7(a--f) of \cite{SI} measured at various constant cooling fields, while varying the measuring field range. Here, while lowering the measuring field, we observe that $\le$50~kOe cooling field is not sufficient to realign the frozen spins even at 5~K temperature and therefore the M--H loops are shifted towards the -ve H-axis and +ve M-axis. However, a measurement field of up to $\pm$90~kOe is sufficient to realign the frozen spins. The parameters obtained from these M--H loops are summarised in Table S3 of \cite{SI}.

\begin{figure}
\includegraphics[width=3.1in]{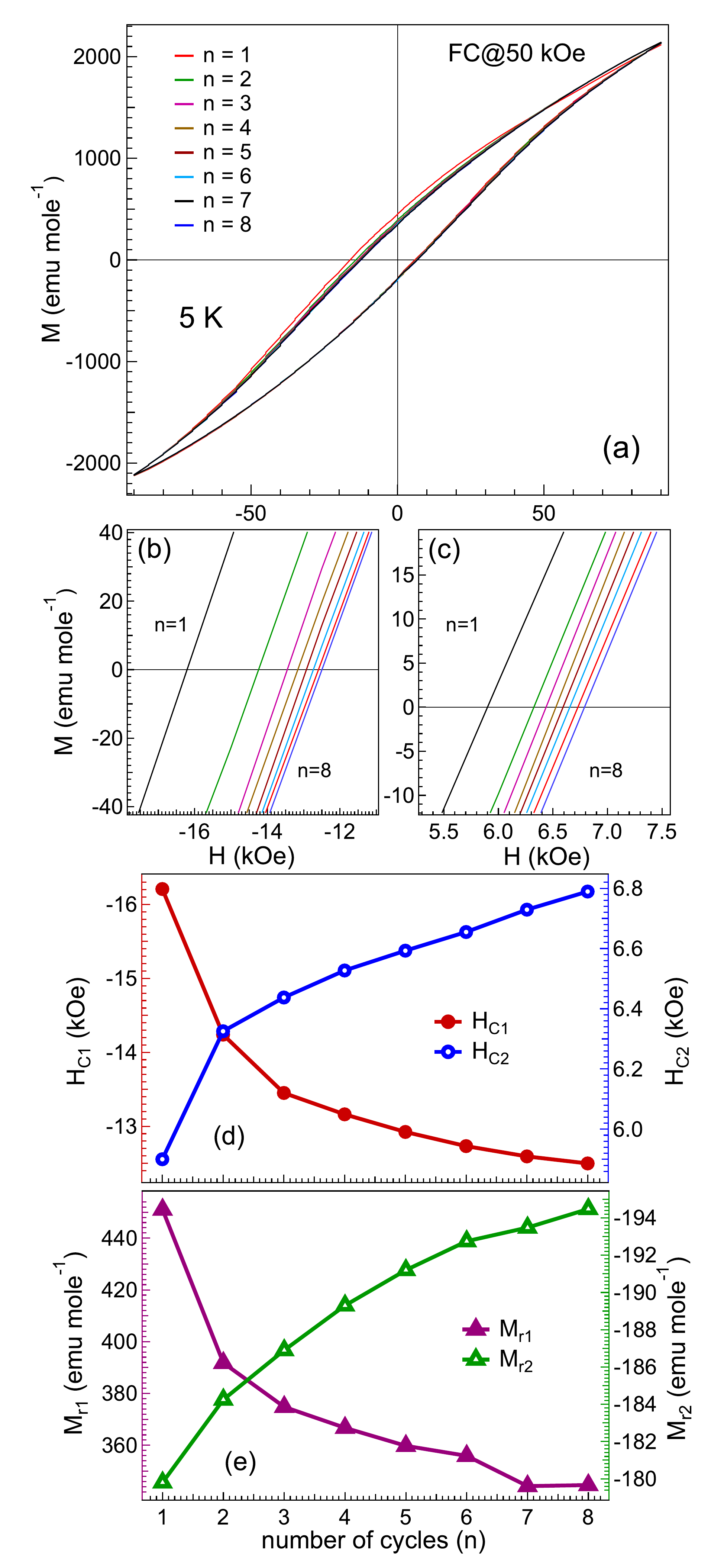}
\caption{(a) The isothermal M--H loops recorded up to consecutive eight ($n=$ 1--8) cycles within $\pm$90~kOe at 5~K under the FC protocol with a cooling field of 50~kOe, an enlarged view of the decay in the (b) H$\rm_{C1}$ (-ve branch) and (c) H$\rm_{C1}$ (+ve branch), the variation in the (d) H$\rm_{C1}$ (-ve branch) on the left-axis and H$\rm_{C2}$ (+ve branch) on the right-axis, and (e) the M$\rm_{r1}$ (+ve branch) on the left-axis and M$\rm_{r2}$ (-ve branch) on the right-axis with the number of subsequent cycles.}
\label{fig:MTE1}
\end{figure}

In order to highlight the observation of an unconventional exchange bias behavior we measure the isothermal M--H loops at 5~K within $\pm$90~kOe measuring field by varying the cooling field in the range of 10--90~kOe, as shown in Fig.~S8(a) of \cite{SI}. The zoomed view in Fig.~S8(b) of \cite{SI} highlights the variation of the +ve and -ve branches of the coercive field, where we can clearly see that the +ve branch of coercive field exhibits a smaller variation with the cooling field while the -ve branch of coercive field changes significantly. We observe that for the cooling fields up to 50~kOe, the value of the coercive field increases, i.e., the -ve branch of the M-H loop intercepts the H-axis increases (marked with the arrow); while, for the cooling fields beyond 50~kOe, a decrease in the value of intercept on the -ve H-axis is observed up to a cooling field of 90~kOe, as highlighted by arrows in Fig.~S8(c) of \cite{SI}. At large cooling field, the size of the FM cluster grows, and therefore a smaller measuring field is enough to realign the frozen spins which results in a decrease in the exchange bias. The extracted parameters are summarized in Table S4 of \cite{SI}. 

It is important to study the magnetic training effect (MTE), which helps to understand the intrinsic exchange bias where the values of H$\rm_C$ and H$\rm_{EB}$ are expected to exhibit a dramatic reduction as the sample is continuously field cycled at a fixed temperature below T$\rm_f$ \cite{MalozenmoffPRB87, MalozenmoffJAP88}. Therefore, eight consecutive isothermal M--H loops were measured at 5~K after a FC protocol with a cooling field of 50~kOe, as shown in Fig.~\ref{fig:MTE1}(a). Further, we highlight the shift in the M--H loop branches on the -ve and +ve H-axis in Figs.~\ref{fig:MTE1}(b, c), respectively. We find an abrupt drop in the values of H$\rm_C$ and H$\rm_{EB}$ from cycle 1 to cycle 2, and it further decreases when the M--H loops are measured in the consecutive cycling. The MTE help to understand the effect of magnetization due to the FM and AFM interface in the sample and a decrease in the coercive fields (H$_{\rm C}$) and remanence magnetization (M$_{\rm r}$) with the consecutive M--H loops manifest that frozen spins relax at the interface \cite{HochstratPRB02}. We plot the values of the H$\rm_{C1}$ (-ve branch) on the left-axis (solid red circles) and H$\rm_{C2}$ (+ve branch) on the right-axis (open blue circles) in Fig.~\ref{fig:MTE1}(d), and M$\rm_{r1}$ (+ve branch) on the left-axis (solid purple triangles) and M$\rm_{r2}$ (-ve branch) on the right-axis (open green triangles) in Fig.~\ref{fig:MTE1}(e) as a function of number of cycles/loop index ($n$). We find that the values on -ve H-axis (H$\rm_{C1}$) change from -16.2~kOe ($n=$ 1) to -12.5~kOe ($n=$ 8); whereas on the +ve H-axis (H$\rm_{C2}$), the change is from 5.9~kOe ($n=$ 1) to 6.8~kOe ($n=$ 8), see Fig.~\ref{fig:MTE1}(d). Like wise, the values of M$\rm_{r1}$ changes from 450~emu/mole ($n=$ 1) to 300~emu/mole ($n=$ 8); whereas the M$\rm_{r2}$ changes from -180~emu/mole ($n=$ 1) to -194~emu/mole ($n=$ 8), see Fig.~\ref{fig:MTE1}(e).  

\begin{figure}
\includegraphics[width=2.9in]{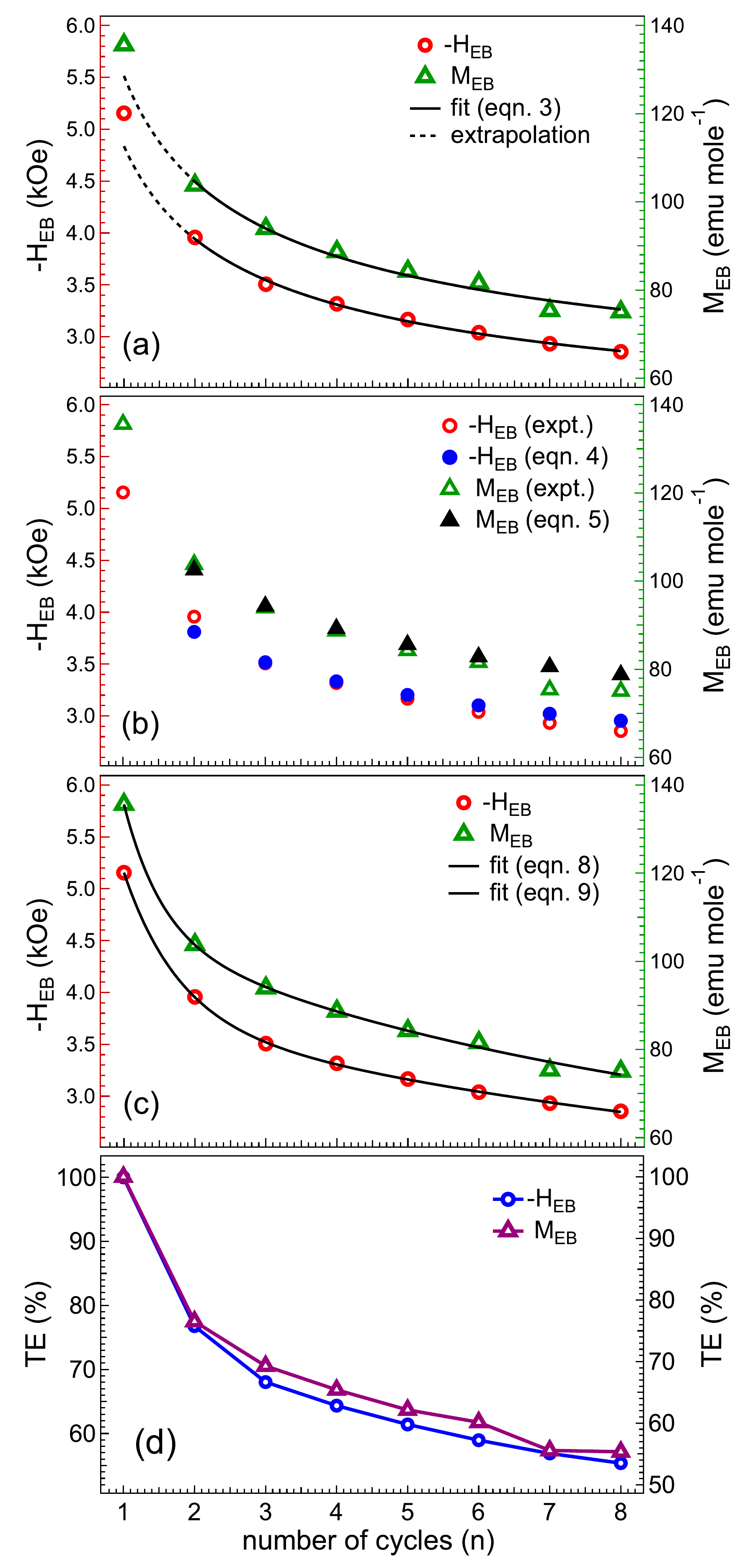}
\caption{(a) The variation of exchange bias field (H$\rm_{EB}$, open red circles on left-axis) and magnetization (M$\rm_{EB}$, open green triangles on right-axis) with the number of cycles ($n$). The power-law fit (solid black lines) to H$\rm_{EB}$ and M$\rm_{EB}$ for $n>$1 using the equations~3, respectively, and the dotted black line extrapolates the same up to $n=$1, (b) the experimental (open symbols) and theoretically generated (solid symbols) values of H$\rm_{EB}$ and M$\rm_{EB}$ using equations~4 and 5, respectively, (c) the best fits (solid black lines) to the dependence of H$\rm_{EB}$ and M$\rm_{EB}$ with $n$ using equations~8 and 9, respectively, and (d) the relative reduction of the H$\rm_{EB}$ and M$\rm_{EB}$ values with $n$ according to the equations~10 and 11, respectively.}
\label{fig:MTE2}
\end{figure}

Moreover, to understand the MTE quantitatively we follow the idea of Binek $et~al.$ that the training of antiferromagnetic spins at the FM--AFM interface gives rise to the reorientation of the domains, i.e., $n$ dependence of exchange bias parameters (H$\rm_{EB}$/M$\rm_{EB}$) \cite{BinekPRB04}. Therefore, a power law fit is used to approximate the arrangement of spins at the interface and can be simulated by the following equations for $n>$ 1 \cite{BinekPRB04},
\begin{eqnarray}
\rm H^n_{EB} - H^{\infty}_{EB} &=& \frac{k_H}{\sqrt{n}}~~~~ \rm and~~~~ \rm M^n_{EB} - M^{\infty}_{EB} = \it \frac{k_M}{\sqrt{n}} 
\end{eqnarray}
where H$\rm^n_{EB}$ (M$\rm^n_{EB}$) and H$\rm^{\infty}_{EB}$ (M$\rm^{\infty}_{EB}$) are the EB fields (magnetization) in the n$\rm^{th}$ cycle and the limit of infinite loops, respectively; and k$_H$, k$_M$ are the system-dependent constants corresponding to the EB field and magnetization, respectively; and $n$ is the number of cycles. The best fits to the calculated -H$\rm_{EB}$ (solid black line) and M$\rm_{EB}$ (solid black line) using the power law equations~3 for $n>$ 1 are presented in Fig.~\ref{fig:MTE2}(a) and the obtained fitting parameters are summarized in Table S4 of \cite{SI}. Note that the extrapolation of the power law fit up to $n=$ 1 underestimates the values of H$\rm_{EB}$ (-4.84~kOe as compared to -5.15~kOe) and M$\rm_{EB}$ (128.6~emu mole$^{-1}$ as compared to 135.6~emu mole$^{-1}$). This difference in the values is correlated to the abrupt drop in the -H$\rm_{EB}$ and M$\rm_{EB}$ values from $n=$ 1 to $n=$ 2, while for $n$$\ge$2 the drop is monotonous and therefore follow the power law precisely to the experimental data for $n>$ 1. This behavior is correlated to the spin relaxation at the FM/AFM interface while field-switching and a uniform decrease is related to the thermally activated spin-rearrangement at the magnetically disordered FM/AFM interface \cite{MishraPRL09}. 

To get the exchange bias, the large AFM spin correlation is necessary to pin the FM layer in the process of magnetization reversal. The relaxation of the spins to the equilibrium because of the applied magnetic field scales to the third power of order parameter and therefore, the non-exponential relaxation of the FM spins below T$\rm_f$ can be approximated with the equations given below \cite{BinekPRB04};
\begin{eqnarray}
\rm H_{EB}(n+1) &=& \rm H_{EB}(n) - \gamma'_H\left[(H_{EB}(n) - H^{\infty}_{EB})\right]^3\\
\rm M_{EB}(n+1) &=& \rm M_{EB}(n) - \gamma'_M\left[(M_{EB}(n) - M^{\infty}_{EB})\right]^3
\end{eqnarray}
where H$\rm_{EB}$($n$) and M$\rm_{EB}$($n$); H$\rm_{EB}$($n+$1) and M$\rm_{EB}$($n+$1)); and H$\rm^{\infty}_{EB}$ and H$\rm^{\infty}_{EB}$ are the EB fields and magnetization in the n$\rm^{th}$, (n+1)$\rm^{th}$ cycle and in the limit of infinite loops, respectively. Here, the $\gamma'_H$ and $\gamma'_M$ are the system dependent constant and can be calculated from the experimental data using the following recursive relations \cite{BinekPRB04},
\begin{eqnarray}
\rm \gamma'_H = \frac{1}{N-1} \sum_{n=2}^{N} \frac{\left[H_{EB}(n) - H_{EB}(n+1)\right]}{\left[H_{EB}(n) - H_{EB}^{\infty}\right]^3}\\
\rm \gamma'_M = \frac{1}{N-1} \sum_{n=2}^{N} \frac{\left[M_{EB}(n) - M_{EB}(n+1)\right]}{\left[M_{EB}(n) - M_{EB}^{\infty}\right]^3}
\end{eqnarray}
where the value of H$\rm_{EB}^{\infty}$ (M$\rm_{EB}^{\infty}$) is utilized as additional input from the power-law fit [equation 3]. The optimized values of $\gamma'_H$ and $\gamma'_M$ according to the equation~3 are included in Table S4 of \cite{SI}. The H$\rm_{EB}$ and M$\rm_{EB}$ values are theoretically generated for $n>$ 1 using the above equations and superimposed to the experimental data in Fig.~\ref{fig:MTE2}(b), where we find that on the left-axis (right-axis), experimentally and theoretically generated H$\rm_{EB}$ (M$\rm_{EB}$) are shown in open red circles (open green triangles) and closed blue circles (closed black triangles), respectively. The generated values of H$\rm_{EB}$ and M$\rm_{EB}$ are in good agreement with the experimental data except for a small deviation for the higher values of $n$. 

Since the presence of spin frustration at the FM/AFM interface modify the magnetic anisotropy leads to the two different species of uncompensated spins, termed as the frozen and rotatable spins \cite{MishraPRL09}, which are strongly exchange coupled to the AFM and FM layers, respectively. This idea of two types of uncompensated spin species can be understood by a quantitative model proposed for the relaxation/decay of the exchange bias with number of cycles where the EB field can be estimated using the equations below \cite{MishraPRL09}, 
\begin{eqnarray}
\rm H^n_{EB} = \rm H^{\infty}_{EB} + A_f exp\left[\frac{-n}{P_f}\right] + A_r exp\left[\frac{-n}{P_r}\right]\\
\rm M^n_{EB} = \rm M^{\infty}_{EB} + A_f exp\left[\frac{-n}{P_f}\right] + A_r exp\left[\frac{-n}{P_r}\right]
\end{eqnarray}
where A$\rm_f$ and P$\rm_f$ are the parameters related to the change in frozen spins, while A$\rm_r$ and P$\rm_r$ are the evolving parameters related to the rotatable spins. Here, the parameters P$\rm_f$ and P$\rm_r$ are dimensionless, while the A$\rm_f$ and A$\rm_r$ have the dimensions of magnetic field and magnetization, respectively. The best fits (solid black line) using the equations~8 and 9 for the H$\rm_{EB}$ (on the left-axis) and M$\rm_{EB}$ (on the right-axis), respectively, are presented in Fig.~\ref{fig:MTE2}(c) and the output parameters are summarized in Table S4 of \cite{SI}. Here, we find that the major contribution in the exchange bias appears to be dominated by the uncompensated spins which are rotating at the interface (rotatable spins) as compared to the frozen spins at the FM/AFM interface. Since the P$\rm_r$ and P$\rm_f$ have a ratio of nearly 13:1 (for H$\rm_{EB}$) and ~14:1 (for M$\rm_{EB}$), which manifests that the frozen spins relax slowly as compared to the rotatable spins, and therefore it indicates that the rotatable spins governs the magnetic training effect observed in the sample with the number of cycles. 

\begin{figure*}
\includegraphics[width=6.2in]{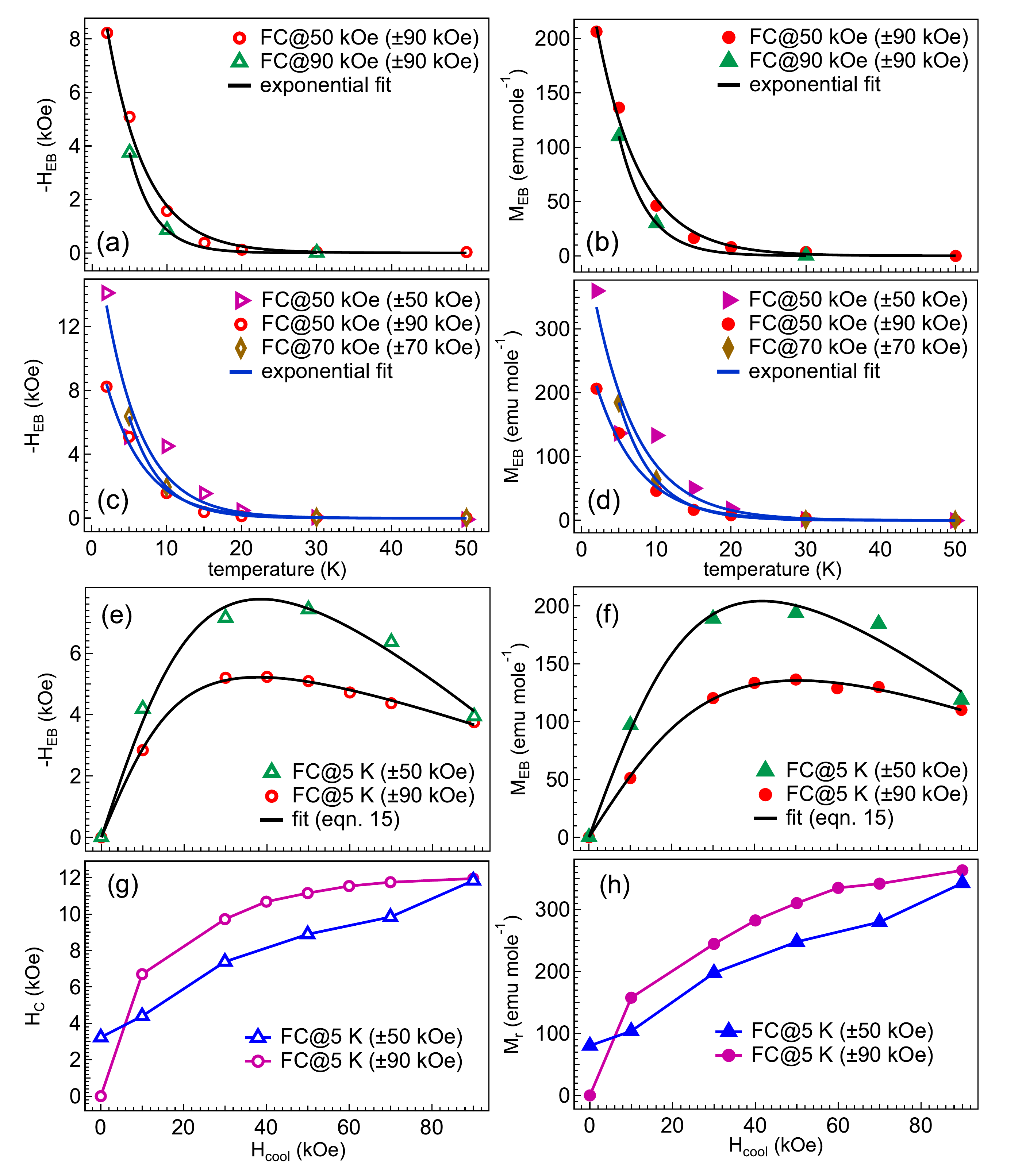}
\caption{The best exponential fits (solid black lines) to the exchange bias (a, c) field (H$\rm_{EB}$), and (b, d) magnetization (M$\rm_{EB}$) in 2--50~K range and comparing the effect of cooling fields (50~kOe, 70~kOe, 90kOe) as well as the measuring field range within $\pm$50~kOe, $\pm$70~kOe and $\pm$90~kOe. The cooling field dependence of the (e) H$\rm_{EB}$ and (f) M$\rm_{EB}$ at 5~K with the best fits (solid black lines) to the data using equation~15, and the change in (g) H$\rm_{C}$ and (h) M$\rm_{r}$ values with the cooling field, measured at 5~K.}
\label{fig:EB_Fit}
\end{figure*}

Furthermore, the relaxation/decay in the EB field (magnetization) with the number of cycles can also be affected due to the spin rearrangement at the interface. The relative percentage of the reduction/decay in the EB field [TE$\rm_H$(\%)] and the magnetization [TE$\rm_M$(\%)] can be estimated using the equations below \cite{VenturaPRB08},
\begin{eqnarray}
\rm TE_H(\%) &=& \rm\left[1 - \frac{(H^1_{EB} - H^n_{EB})}{H^1_{EB}}\right]\times 100\%\\
\rm TE_M(\%) &=& \rm\left[1 - \frac{(M^1_{EB} - M^n_{EB})}{M^1_{EB}}\right]\times 100\%
\end{eqnarray}
where, the H$\rm^1_{EB}$ (M$\rm^1_{EB}$) and H$\rm^n_{EB}$ (M$\rm^n_{EB}$) are the exchange bias field (magnetization) at first and n$\rm^{th}$ cycles. We plot the TE$\rm_H$(\%) and TE$\rm_M$(\%) in Fig.~\ref{fig:MTE2}(d) on the left-axis (open blue circles) and right-axis (open purple triangles) using equations~10 and 11, respectively. Here, we observe that the exchange bias field (magnetization) decreases very sharply to around 77\% in the initial two cycles, and after that this decay rate reduces, which is consistent with the reduction in the amplitude of spin-rearrangement happening with the number of cycles. 

Now, we perform quantitative analysis to understand the origin of the exchange bias  where we find above that there is a logarithmic decrease of H$\rm_{EB}$ and M$\rm_{EB}$ in 2--50~K range for the cooling fields of 90~kOe, 70~kOe, and 50~kOe, see Figs.~3(b, c, e). We model this variation in H$\rm_{EB}$ and M$\rm_{EB}$ using the following exponential relations \cite{MoutisPRB21},
\begin{eqnarray}
\rm H_{EB} &=& \rm H_{EB}^0~exp(-T/T_0)\\
\rm M_{EB} &=& \rm M_{EB}^0~exp(-T/T_1)
\end{eqnarray}
where the T$_0$ (T$_1$) and the H$\rm_{EB}^0$ (M$\rm_{EB}^0$) are constants and the extrapolation of H$\rm_{EB}$ (M$\rm_{EB}$) to the absolute zero temperature, respectively. The best fits (solid black lines) to the experimental H$\rm_{EB}$ and M$\rm_{EB}$ data for the cooling fields of 50~kOe and 90~kOe for the measuring fields of $\pm$90~kOe are presented in Fig.~\ref{fig:EB_Fit}(a, b), respectively, and the output parameters are summarized in Table S6 of \cite{SI}. Interestingly, an exponential dependence of the exchange bias parameters originates from the competitive/coupled FM--AFM interactions in the sample where the SE and DE interactions are responsible for the presence of AFM and FM interactions, respectively \cite{MoutisPRB21, HuangPRB08}. Also, we compare the exponential decay of the exchange bias parameters H$\rm_{EB}$ and M$\rm_{EB}$ with the different cooling (50~kOe and 70~kOe) and measuring ($\pm50$~kOe, $\pm$90~kOe$, \pm$70~kOe) fields in Figs.~\ref{fig:EB_Fit}(c, d), respectively along with the exponential best fits (solid blue lines). The estimated exchange bias parameters H$\rm_{EB}^0$ and M$\rm_{EB}^0$ are found to decrease for lower values of the cooling fields, i.e., for 50~kOe (12.4~kOe and 299~emu/mole) as compared to 90~kOe (16.35~kOe and 403~emu/mole) for the fixed measuring field of $\pm$90~kOe. However, this effect is variable for the different values of measuring fields, i.e., for a higher value of measuring field $\pm$90~kOe, the values of H$\rm_{EB}^0$ and M$\rm_{EB}^0$ decrease stronger as compared to $\pm$70~kOe and $\pm$50~kOe (see Table S6 of \cite{SI}). Further, we present the dependence of the exchange bias field (H$\rm_{EB}$) and magnetization (M$\rm_{EB}$) with the cooling fields (0--90~kOe) at 5~K in Figs.~\ref{fig:EB_Fit}(e, f), respectively. Here, we find that the H$\rm_{EB}$ and M$\rm_{EB}$ increases up to 50~kOe cooling field and then start decreasing monotonously. This behavior is contrary to the conventional saturating behavior of H$\rm_{EB}$ and M$\rm_{EB}$ with the cooling fields, as observed in \cite{AnusreePRB20, PramanikPRB21}. However, note that similar unconventional EB behavior is also reported in the cluster spin glass compounds \cite{KumarPRB21, GiriJAP12}. Therefore, this observation manifests that the exchange bias appearing due to the cooling fields up to 50~kOe enhances competitive FM--AFM interactions and then after $>$50~kOe starts suppressing it (discussed below in detail). The decrease in the H$\rm_{EB}$ and M$\rm_{EB}$ is associated with the increase in the values of H$\rm_{C}$ and M$\rm_{r}$, see Figs.~\ref{fig:EB_Fit}(g, h), respectively, which also suggests the growth of FM clusters and/or alignment of the FM clusters in the magnetic field direction for the higher values of cooling fields. However, this increase in H$\rm_{C}$ and M$\rm_{r}$ is more prominent for the measuring field range of $\pm$90~kOe as compared to $\pm$50~kOe, which also supports the hypothesis of more growth/alignment of FM clusters in the larger measuring field range. 

Notably the decrease in H$\rm_{EB}$ and M$\rm_{EB}$ values for the cooling field $>$50~kOe is found to be invariant of measuring field ranges, i.e., $\pm$50~kOe and/or $\pm$90~kOe [see Figs.~\ref{fig:EB_Fit}(e, f)], which confirms the intrinsic nature of the observed unconventional behavior. The H$\rm_{EB}$ values can be influenced to some extent by the increase/growth in the size of the FM cluster and/or alignment of the degree of the FM cluster at particular cooling field. The cooling field strongly affects the orientation of the frozen spins and governs the competitive/coupled FM--AFM interaction at the interfaces in the CSG phase. According to the Bean's relation for the FM--AFM interface, the exchange bias field is inversely proportional to the thickness of the FM layer (t$\rm_{FM}$), as per the equation below \cite{BeanPR56, KarmakarPRB08}, 
\begin{eqnarray}
\rm -H_{EB}=J_{ex}\frac{S_{AFM}S_{FM}}{\mu_0t_{FM}M_{FM}}
\end{eqnarray}
where J$\rm_{ex}$ is the exchange integral across the FM/AFM interface per unit area, S$\rm_{FM}$ and S$\rm_{AFM}$ are the interface ferromagnet and antiferromagnet magnetization, respectively, and M$\rm_{FM}$ is the magnetization of the FM layer. Since, an increase in the cooling field results in the growth of the FM cluster size and degree of alignment, which advocates the increase in the values of t$\rm_{FM}$ and M$\rm_{FM}$, respectively. Also, higher cooling field enhances the values of interfacial magnetization (S$\rm_{FM}$ and S$\rm_{AFM}$) at the FM/AFM interface in the CSG phase \cite{KarmakarPRB08}. Thus, as per the above equation~14, the H$\rm_{EB}$ increases with the cooling field. Although, at higher values the size of FM clusters further grows and despite an increase in the values of t$\rm_{FM}$ and M$\rm_{FM}$ with the cooling field, a decrease in the FM/AFM interface area reduces S$\rm_{FM}$ and S$\rm_{AFM}$. Therefore, owing to the increased value of the denominator and a reduction in the value of the numerator of equation~14, we find a reduction in the values of H$\rm_{EB}$ at higher values of cooling fields, which is in good agreement with \cite{KarmakarPRB08}. 

In order to further investigate the origin of the unconventional exchange bias observed in the sample, we follow the relation given by Niebieskikwiat and Salamon \cite{NiebieskikwiatPRB05} at low temperatures ($\mu_0$H$_{cool}$ $<$ k$_B$T) below T$\rm_f$, which advocates that owing to the glassy behavior of surface spins, interface moments ($m_i$ $\propto$ H$\rm_{EB}$) remain frozen and quantitative dependence of H$\rm_{EB}$ on the cooling field can be modelled such that \cite{NiebieskikwiatPRB05, KarmakarPRB08}, 
\begin{eqnarray}
\rm -H_{EB}\propto \frac{M_E}{M_S}\propto J_i\left[\frac{J_i\mu_o}{(g\mu_B)^2}L\left(\frac{\mu H_{cool}}{k_BT_f}\right)+H_{cool}\right]
\end{eqnarray}
where $J_i$, $g$, $L(x)$, $k_B$, $\mu_B$, and H$\rm_{cool}$ are the interface exchange coupling constant, gyromagnetic factor, Langevin function, Boltzmann constant, Bohr magneton, and cooling field. Here, $\mu$ is defined as the magnetic moment of FM particles, i.e., $\mu = N_{FM}\mu_0$, where $N_{FM}$ is the number of spins in the FM cluster. In the above equation, for the smaller cooling field the first term dominates, i.e., -H$\rm_{EB}\propto J_i^2$ and for the larger values of the cooling field the second term dominates, i.e., -H$\rm_{EB}\propto J_i$. Therefore, for a negative value of $J_i<$0, the exchange bias parameters first increase with the cooling field owing to the dominance of the first term ($J_i^2$) and then start decreasing or a sign reversal is observed at the large values of the cooling field, where the second term ($J_i$) in equation~15 dominates \cite{KellerPRB02, NoguesPRB96}. We fit the cooling field dependence of H$\rm_{EB}$ and M$\rm_{EB}$ using equation~15 [solid black lines in Figs.~\ref{fig:EB_Fit}(e, f)] where $J_i$ and $N_{FM}$ are considered as free parameters. We obtain the values of $J_i =$ -10.24$\pm$0.22~meV and $N_{FM} =$ 2.4$\pm$0.2 for the measuring field of $\pm$50~kOe and $J_i =$ -12.55$\pm$0.49~meV and $N_{FM} =$ 3.1$\pm$0.1 for the measuring field of $\pm$90~kOe. Here, we find that the interface exchange coupling constant ($J_i$) higher as compared to reported in refs.~\cite{NiebieskikwiatPRB05, KarmakarPRB08}, which is correlated to the observed giant values of exchange bias in the present sample. Interestingly, the negative values of J$_i$ explain the dominance of AFM exchange interaction at the interface of FM/AFM interactions in the CSG region, and smaller value of $N_{FM}$ manifest that the size of FM clusters is small. Therefore, it advocates the larger interaction area between the AFM/FM interface and further supports the observed giant values of exchange bias. Moreover, a higher value of J$_i$ and $N_{FM}$ for the measuring fields $\pm$90~kOe explains the decrease in the values of EB parameters at higher measuring and cooling fields.

\begin{figure}[h]
\includegraphics[width=3.5in]{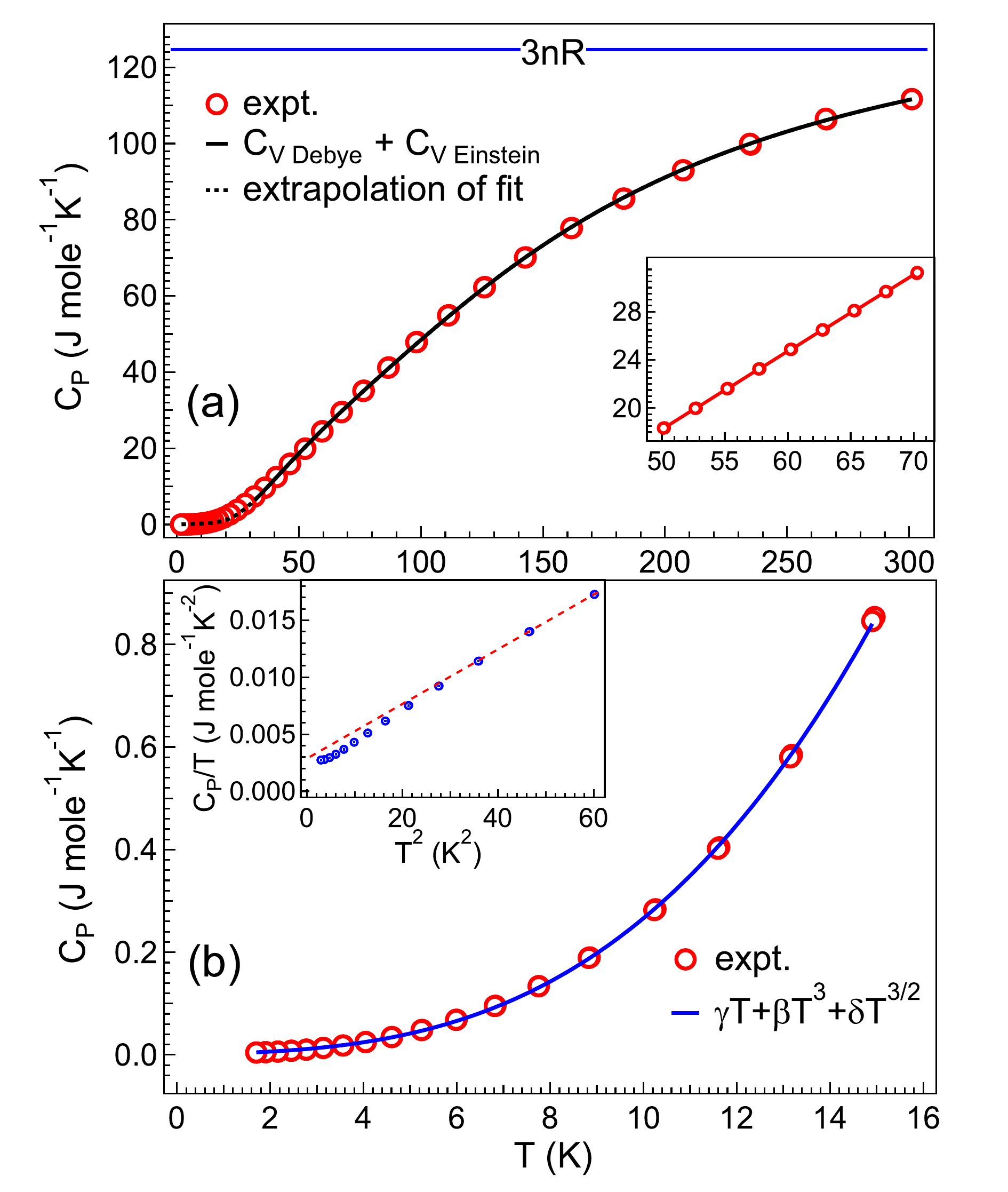}
\caption{(a) The specific heat data (red circles) fitted in the temperature range of 30--300 K (solid black line) using a combination of Debye and Einstein models following the Pad\'e approximation; blue solid line represents the value according to the Dulong-Petit law (3nR), and (b) a best fit (solid blue line) below 15~K including the electronic ($\gamma T$), lattice ($\beta T^3$), and magnetic ($\delta T^{3/2}$) terms, inset highlights the deviation in the C$\rm_P$/T versus T$^2$ from a straight line in 2--8~K range.}
\label{fig:Cp}
\end{figure} 

Finally, we present the zero-field specific heat data [C$\rm_p$(0, T)] in the range of 1.9--300~K in Fig.~\ref{fig:Cp}(a), where the C$\rm_p$ value 111.7~J mole$^{-1}$K$^{-1}$) is found to be less than the Dulong-Petit limit, C$\rm_V$ = 3nR = 124.7~J mole$^{-1}$K$^{-1}$ (solid blue line), where $n$ is the numbers of ions per mole (5 for this sample) and R is the universal gas constant (8.314~J mole$^{-1}$K$^{-1}$), respectively \cite{Kittel05}. We observe that the specific heat value decreases with temperature and there is no anomaly in the measured temperature range, which manifests an absence of any long-range magnetic ordering in the sample. Also, there is no Shottky-like anomaly in the present Nb substituted sample, in contrast to the reports in La$_{1-x}$Sr$_x$CoO$_3$ \cite{HeAPL09, HePRB09}. The observed specific heat is a cumulative effect of contributions from the lattice due to phonons (C$\rm_{lattice}$), electronic because of the conduction electrons (C$\rm_{elect}$), and a magnetic specific heat (C$\rm_{magn}$), which can be written as C$_p$(T) = C$\rm_{lattice}$+ C$\rm_{elect}$+ C$\rm_{magn}$. First, we fit the specific heat data only with the lattice contributions, i.e., from a combination of Debye and Einstein heat-capacity models, using the formula C$\rm_{lattice}$ = mC$\rm_{V(Debye)}$ + (1-m)C$\rm_{V(Einstein)}$. The contribution of the Debye heat-capacity model at a constant volume is given as \cite{Gopal66},
\begin{eqnarray}
C{\rm_{V(Debye)}} (T) = 9nR\left(\frac{T}{\theta_D}\right)^3 \int_{0}^{\theta_D/T} \frac{x^4e^x}{(e^x-1)^2}dx
\end{eqnarray}
where $\theta_D$ is Debye temperature and its origin belongs to the acoustic phonon vibrations in the sample and at a constant volume the contribution of the Einstein heat-capacity model can be written as \cite{Gopal66},
\begin{eqnarray}
C{\rm_{V(Einstein)}} (T) = 3nR\left(\frac{\theta_E}{T}\right)^2 \frac{e^{\theta_E/T}}{(e^{\theta_E/T}-1)^2} 
\end{eqnarray}
where $\theta_E$ is the Einstein temperature corresponding to the optical phonons. In Fig.~\ref{fig:Cp}(a), from the best fitting in 30-300~K range (solid black line) using above equations we obtain a contribution of around 71\% and 29\% from the Debye and Einstein models, respectively. Here, we use the Pad\'e approximation for the Debye model \cite{GoetschPRB12} for fitting the specific heat data. The estimated values of $\theta_D$ and $\theta_E$ are = 688$\pm$10~K and 171$\pm$3~K, respectively. We find that the Nb substitution enhances the value of $\theta_D$ as compared to La$_{1-x}$Sr$_x$CoO$_3$ (400-500~K) \cite{HeAPL09, HePRB09}. This enhancement in $\theta_D$ value can be associated with the non-saturating value of the specific heat at room temperature \cite{AjayPRB20}. The fitted curve is extrapolated to the lowest measured temperature (dashed black line) in Fig.~\ref{fig:Cp}(a), which shows a deviation in the specific heat at low temperatures from the  lattice approximation, which manifests that an additional contribution is required to understanding the low temperature heat capacity behavior. 

Further, we observe that the specific heat exhibits a nonlinear behavior at low temperatures, see the inset in Fig.~\ref{fig:Cp}(b). This indicates that the specific heat does not follow $C_p = \gamma T+\beta T^3$ (where, $\gamma T$ and cubic term $\beta T^3$ correspond to the electronic and lattice contributions, respectively) at low temperatures and that is believed to be due to the short-range magnetic interactions/glassy nature in the sample \cite{AnusreePRB20, AnandPRB12}. Therefore, we fit the C$_p$ data including a magnetic term ($\delta T^{3/2}$) using $C_p = \gamma T+\beta T^3+\delta T^{3/2}$  \cite{AnandPRB12, AnusreePRB20}. The coefficient of the electronic term ($\gamma$) can be utilized to estimate the density of states (DOS) at the Fermi level, i.e., N(E$\rm_F$) as $\gamma = \pi^2k_B^2N(E_F)$/3, the coefficient of lattice contribution ($\beta$) is given by the formula $\beta = (234nk_B)/\theta_D^3$, or, $\theta_D^3 = (12\pi^4nR)/5\beta$ \cite{AnandPRB12, GhivelderPRB99}, and the coefficient of the magnetic term ($\delta$) can be used to estimate the spin-wave stiffness constant ($D$), such that, $\delta = 0.113Ra^3(k_B/D)^{3/2}$, where $a$ is the lattice parameter of elementary perovskite unit cell \cite{Gopal66, GhivelderJMMM98}. The best fit (solid blue line) of the low temperature $C_p$ using the above equation having three terms (electronic+lattice+magnetic) in Fig.~\ref{fig:Cp}(b) provides a good description of the data and the obtained fitting coefficients are $\gamma=$ 2.14$\pm$0.04~mJ mole$^{-1}$-K$^{-2}$, $\beta=$ 0.24$\pm$0.01~mJ mole$^{-1}$K$^{-4}$, and $\delta=$ 0.13$\pm$0.02~mJ mole$^{-1}$K$^{-5/2}$. Here, the estimated value of DOS is found to be N(E$_F$) = 5.47$\times$10$^{23}$~eV$^{-1}$mole$^{-1}$ and the $\theta_D$ value is 323.7~K. This $\theta_D$ value is nearly half of the value estimated from the Debye model due to complex magnetic interactions at low temperatures, which can be attributed to the cluster spin-glass behavior or the excitonic mass enhancement in the low-lying crystal field \cite{AnandPRB12}. Moreover, the estimated value of spin-wave stiffness constant $D$ is 470~meV\AA$^2$, which is in the range of values reported for the perovskite manganites in \cite{GhivelderPRB99, GhivelderJMMM98, WoodfieldPRL97}. \\

\section{\noindent ~Conclusions}

In conclusion, the effect of different external parameters (cooling field, temperature, and measuring fields) on the exchange bias behavior in cluster spin glass (CSG) La$_{0.5}$Sr$_{0.5}$Co$_{0.85}$Nb$_{0.15}$O$_3$ is investigated in detail using the field cooled isothermal M--H loops and magnetic training effect. The magnetic memory effect and aging measurements confirm the presence of a glassy magnetic state. Moreover, the detailed analysis of ac-$\chi$ data confirms the presence of CSG with a characteristic spin-relaxation time-scale of $\tau_0$ = 8.4$\times$10$^{-10}$~s. The observed CSG emerges due to the hole-poor and hole-rich regions owing to the ferromagnetic (FM) double exchange interactions (from Co$^{4+}$-O-Co$^{3+}$) and antiferromagnetic (AFM) super-exchange interactions (from Co$^{3+}$-O-Co$^{3+}$ and Co$^{4+}$-O-Co$^{4+}$), respectively. Interestingly, we find that the small size FM clusters are embedded in the AFM matrix like a conventional FM/AFM interface and therefore results in a giant tunable exchange bias in the sample. The EB parameters (H$\rm_{EB}$ and M$\rm_{EB}$) show unusual dependency on the cooling field. Further, we model this tunable behavior with the interface exchange interaction model and find that the exchange interaction is strong and negative $J_i =$ -10.24$\pm$0.22~meV and $N_{FM} =$ 2.4$\pm$0.2 (measuring field of $\pm$50~kOe) and $J_i =$ -12.55$\pm$0.49~meV and $N_{FM} =$ 3.1$\pm$0.1 (measuring field of $\pm$90~kOe). The specific heat data analysis also provides insights into the complex magnetic interactions at low temperatures. 

\section{\noindent ~Acknowledgements}

RS acknowledges DST, INSPIRE for fellowship. RSD is grateful to the SERB--DST for the financial support through core research grant (project reference no. CRG/2020/003436).


\begin{thebibliography}{99}

\bibitem{MeiklejohnPR56+57} W. H. Meiklejohn and C. P. Bean, New Magnetic Anisotropy, Phys. Rev. {\bf102} 1413 (1956); {\bf105} 904 (1957). 

\bibitem{DienyJMMM94} B. Dieny, Giant magnetoresistance in spin-valve multilayers, J. Magn. Magn. Mater. {\bf136}, 335--359 (1994).

\bibitem{RaduPRB09} F. Radu, M. Etzkorn, R. Siebrecht, T. Schmitte, K. Westerholt, and H. Zabel, Interfacial domain formation during magnetization reversal in exchange-biased CoO/Co bilayers, Phys. Rev. B {\bf67}, 134409 (2009).

\bibitem{TsangIEEE82} C. Tsang and R. Fontana Jr., Fabrication and wafer testing of barber-pole and exchange-biased narrow-track MR sensors, IEEE Trans. Magn. {\bf18}, 1149 (1982).

\bibitem{DienyPRB91} B. Dieny, V. S. Speriosu, S. S. P. Parkin, J. C. Scott, B. A. Gurney, D. R. Wilhoit, and D. Mauri, Giant magnetoresistive in soft ferromagnetic multilayers, Phys. Rev. B {\bf43}, 1297 (1991).

\bibitem{NoguesPRL96} J. Nogu\'es, D. Lederman, T. J. Moran, and I. K. Schuller, Positive exchange bias in FeF$_2$-Fe bilayers, Phys. Rev. Lett. {\bf76}, 4624 (1996).

\bibitem{AliNM07} M. Ali, P. Adie, C. H. Marrows, D. Greig, B. J. Hickey, and R. L. Stamps, Exchange bias using a spin glass, Nat. Mat. {\bf6}, 70 (2007).

\bibitem{SkumryevNature03} V. Skumryev, S. Stoyanov, Y. Zhang, G. Hadjipanayis, D. Givord, and J. Nogu\'es, Beating the superparamagnetic limit with exchange bias, Nature {\bf423}, 850 (2003).

\bibitem{VelthuisAPL99} S. G. E. te Velthuis, G. P. Felcher, J. S. Jiang, A. Inomata, C. S. Nelson, A. Berger, and S. D. Bader, Magnetic configurations in exchange-biased double superlattices, Appl. Phys. Lett. {\bf75}, 4174 (1999).

\bibitem{PengPRB00} D. L. Peng, K. Sumiyama, T. Hihara, S. Yamamuro, and T. J. Konno, Magnetic properties of monodispersed Co/CoO clusters, Phys. Rev. B {\bf61}, 3103 (2000).

\bibitem{StampsJPD00} R. L. Stamps, Mechanisms for exchange bias, J. Phys. D: Appl. Phys. {\bf33}, R247 (2000).

\bibitem{ManivNP21} E. Maniv, R. A. Murphy, S. C. Haley, S. Doyle, C. John, A. Maniv, S. K. Ramakrishna, Y.-L. Tang, P. Ercius, R. Ramesh, A. P. Reyes, J. R. Long, and J. G. Analytis, Exchange bias due to coupling between coexisting antiferromagnetic and spin-glass orders, Nat. Phys. {\bf17}, 525 (2021).

\bibitem{DingPRB13} J. F. Ding, O. I. Lebedev, S. Turner, Y. F. Tian, W. J. Hu, J. W. Seo, C. Panagopoulos, W. Prellier, G. Van Tendeloo, and T. Wu, Interfacial spin glass state and exchange bias in manganite bilayers with competing magnetic orders, Phys. Rev. B {\bf87}, 054428 (2013).

\bibitem{KarmakarPRB08} S. Karmakar, S. Taran, E. Bose, B. K. Chaudhuri, C. P. Sun, C. L. Huang, and H. D. Yang, Evidence of intrinsic exchange bias and its origin in spin-glass-like disordered L$_{0.5}$Sr$_{0.5}$MnO$_3$ manganites (L = Y, Y$_{0.5}$Sm$_{0.5}$, and Y$_{0.5}$La$_{0.5}$), Phys. Rev. B {\bf77}, 144409 (2008).

\bibitem{AntelPRL99} W. J. Antel Jr., F. Perjeru, and G. R. Harp, Spin Structure at the Interface of Exchange Biased FeMn/Co Bilayers, Phys. Rev. Lett. {\bf83}, 1439 (1999).

\bibitem{OhldagPRL03} H. Ohldag, A. Scholl, F. Nolting, E. Arenholz, S. Maat, A. T. Young, M. Carey, and J. St$\rm\ddot{o}$hr, Correlation between Exchange Bias and Pinned Interfacial Spins, Phys. Rev. Lett. {\bf}, 017203 (2003).

\bibitem{GruytersPRB00} M. Gruyters and D. Riegel, Strong Exchange bias by a single layer of independent antiferromagnetic grains: The CoO/Co model system, Phys. Rev. B {\bf63}, 052401 (2000).

\bibitem{Ajay_PRB24} Ajay Kumar, B. Schwarz, and R. S. Dhaka, Correlation between exchange bias and antisite disorder in Sr$_{2-x}$La$_x$CoNbO$_6$ ($x=$ 0, 0.2), Phys. Rev. B {\bf109}, 104434 (2024).

\bibitem{GiriJPCM11} S. Giri, M. Patra, and S. Majumdar, Exchange bias effect in alloys and compounds, J. Phys.: Condens. Matter {\bf23}, 072301 (2011).

\bibitem{WangPRB04} H. Wang, T. Zhu, K. Zhao, W. N. Wang, C. S. Wang, Y. J. Wang, and W. S. Khan, Surface spin glass and exchange bias in Fe$_3$O$_4$ nanoparticles compacted under high pressure, Phys. Rev. B {\bf70}, 092409 (2004).

\bibitem{ZhangAPL04} R. K. Zheng, H. Liu, X. X. Zhang, V. A. L. Roy, and A. B. Djuri$\rm\check{s}$i\'c, Exchange bias and the origin of magnetism in Mn-doped ZnO tetrapods, Appl. Phys. Lett. {\bf85}, 2589 (2004).

\bibitem{FisherPRB88} D. S. Fisher and D. A. Huse, Nonetluilibrium dynamics of spin glasses, Phys. Rev. B {\bf38}, 373 (1988); Equilibrium behavior of the spin-glass ordered phase {\bf38}, 386 (1988).

\bibitem{AnderssonPRB93} J. O. Andersson, J. Mattsson, and P. Nordblad, Overlap length in a Cu-Mn spin glass probed by ac susceptibility, Phys. Rev. B {\bf48}, 13977 (1993).

\bibitem{LeflochEL92} F. Lefloch, J. Hammann, M. Ocio, and E. Vincent, Can Aging Phenomena Discriminate Between the Droplet Model and a Hierarchical Description in Spin Glasses?, Europhys. Lett. {\bf18}, 647 (1992).

\bibitem{LiaoAPR23} X. Liao, S. Wei, Y. Wang, D. Wang, K. Wu, H. Liang, S. Yang, P. Svedlindh, and Y.-J. Zeng, Large Exchange Bias Triggered by Transition Zone of Spin Glass, Adv. Phys. Res. {\bf2}, 2200043 (2023).

\bibitem{MeirzadehACSCS2021} E. Meirzadeh, S. Y. Han, and X. Roy, Exchange bias from frustrated spins, ACS Cent. Sci. {\bf7(8)}, 1295 (2021). 

\bibitem{SunNL12} X. Sun, N. F. Huls, A. Sigdel, and S. Sun, Tuning exchange bias in Core/Shell FeO/Fe$_3$O$_4$ nanoparticles, Nano Lett. {\bf12(1)}, 246 (2012). 

\bibitem{SanthoshPRB18} R. R. Das, P. Parida, A. K. Bera, T. Chatterji, B. R. K. Nanda, and P. N. Santhosh, Giant exchange bias in the single-layered Ruddlesden-Popper perovskite SrLaCo$_{0.5}$Mn$_{0.5}$O$_4$, Phys. Rev. B {\bf98}, 184417 (2018).

\bibitem{CourtimPRB16} L. T. Courtim, E. M. Bittar, F. Stavale, F. Garcia, E. Baggio-Saitovitch, A. Abbate, R. J. O. Mossanek, H. P. Martins, D. Tobia, P. G. Pagliuso, and L. Bufaical, Compensation temperatures and exchange bias in La$_{1.5}$Ca$_{0.5}$CoIrO$_6$, Phys. Rev. B {\bf93}, 174406 (2016).

\bibitem{PrajapatJMMM19} M. Prajapata, S. Marik, K. S. Kumar, D. Singh, R. P. Singh, D. S. Rana, and V. Shelke, Giant room temperature exchange bias effect in "314-type" cobaltate SrCo$_{1-x}$V$_{x}$O$_{3-\delta}$ ($x =$ 0.05, 0.1), J. Magn. Magn. Mater. {\bf478}, 247 (2019).

\bibitem{AngJAP08} R. Ang, Y. P. Sun, X. Luo, C. Y. Hao, X. B. Zhu, and W. H. Song, Exchange bias in the layered cobaltite Sr$_{1.5}$Pr$_{0.5}$CoO$_4$, J. Appl. Phys. {\bf104}, 023914 (2008). 

\bibitem{SahooPRB19} R. C. Sahoo, Y. Takeuchi, A. Ohtomo, and Z. Hossain, Exchange bias and spin glass states driven by antisite disorder in the double perovskite compound LaSrCoFeO$_6$, Phys. Rev. B {\bf100}, 214436 (2019).

\bibitem{SinghJPCM22} P. Singh, R. K. Singh, S. Dixit, N. Patel, M. Alam, S. Dan, A. Jain, K. Anand, V. K. Gangwar, and R. Singh, Double glassy states and large spontaneous and conventional exchange bias in La$_{1.5}$Ca$_{0.5}$CoFeO$_6$ ferrimagnetic double perovskite, J. Phys.: Condens. Matter {\bf34}, 375803 (2022). 

\bibitem{GiriJPD16} S. K. Giri, R. C. Sahoo, P. Dasgupta, A. Poddar, and T. K. Nath, Giant spontaneous exchange bias effect in Sm$_{1.5}$Ca$_{0.5}$CoMnO$_6$ perovskite, J. Phys. D: Appl. Phys. {\bf49}, 165002 {2016}. 

\bibitem{AnusreePRB20} V. K. Anusree, R. R. Das, P. N. Lekshmi, R. Dhal, C. V. Colin, and P. N. Santhosh, Giant exchange bias effect in Ruddlesden-Popper oxides SrLaFe$_{0.25+x}$Mn$_{0.25}$Co$_{0.5-x}$O$_4$ ($x =$ 0, 0.25): Role of the cluster glass magnetic phase in a quasi-two-dimensional perovskite, Phys. Rev. B {\bf102}, 134405 (2020).

\bibitem{GolovanovPRB96} V. Golovanov, L. Mihaly and A. R. Moodenbaugh, Magnetoresistance in La$_{1-x}$Sr$_x$CoO$_3$ for 0.05$\le$x$\le$0.2, Phys. Rev. B {\bf53}, 8207 (1996).

\bibitem{TangPRB06} Y.-k. Tang, Y. Sun, and Z.-h. Cheng, Exchange bias associated with phase separation in the perovskite cobaltite La$_{1-x}$Sr$_x$CoO$_3$, Phys. Rev. B {\bf73}, 174419 (2006).

\bibitem{WuPRB05} J. Wu, J. W. Lynn, C. J. Glinka, J. Burley, H. Zheng, J. F. Mitchell, and C. Leighton, Intergranular Giant Magnetoresistance in a Spontaneously Phase Separated Perovskite Oxide, Phys. Rev. Lett. {\bf94}, 037201 (2005).

\bibitem{ShuklaPRB23} R. Shukla and R. S. Dhaka, Evolution of complex magnetic and transport behavior of La$_{0.5}$Sr$_{0.5}$Co$_{1-x}$Nb$_x$O$_3$, Phys. Rev. B {\bf107}, 165108 (2023).

\bibitem{ShuklaPRB18} R. Shukla and R. S. Dhaka, Anomalous magnetic and spin glass behavior in Nb-substituted LaCo$_{1-x}$Nb$_x$O$_{3}$, Phys. Rev. B {\bf97}, 024430(1--9) (2018).

\bibitem{ZenerPR51_1} C. Zener, Interaction Between the $d$ Shells in the Transition Metals, Phys. Rev. {\bf81}, 440 (1951).

\bibitem{ZenerPR51_2} C. Zener, Interaction between the $d$-Shells in the Transition Metals. II. Ferromagnetic Compounds of Manganese with Perovskite Structure, Phys. Rev. {\bf82}, 403 (1951).

\bibitem{AndersonPR55} P. W. Anderson and H. Hasegawa, Considerations on Double Exchange, Phys. Rev. {\bf100}, 675 (1955).

\bibitem{KuhnsPRL03} P. L. Kuhns, M. J. R. Hoch, W. G. Moulton, A. P. Reyes, J. Wu, and C. Leighton, Magnetic Phase Separation in La$_{1-x}$Sr$_x$CoO$_3$ by $^{59}Co$ Nuclear Magnetic Resonance, Phys. Rev. Lett. {\bf91}, 127202 (2003).

\bibitem{HochPRB04} M. J. R. Hoch, P. L. Kuhns, W. G. Moulton, A. P. Reyes, J. Wu, and C. Leighton, Spin dynamics in La$_{1-x}$Sr$_x$CoO$_3$, Phys. Rev. B {\bf69}, 014425 (2004).

\bibitem{HoJMMM19} T. A. Ho, P. T. Long, N. V. Quang, S. L. Cho, and S. C. Yu, Short and long-range ordering in La$_{1-x}$Sr$_x$CoO$_3$ cobaltites, J. Magn. Magn. Mater. {\bf477}, 396--403 (2019).

\bibitem{BagPRB18} P. Bag, P. R. Baral, and R. Nath, Cluster spin-glass behavior and memory effect in Cr$_{0.5}$Fe$_{0.5}$Ga, Phys. Rev. B {\bf98}, 144436 (2018).

\bibitem{Kumar_PRB1_20} A. Kumar and R. S. Dhaka, Unraveling magnetic interactions and the spin state in insulating Sr$_{2-x}$La$_x$CoNbO$_6$, Phys. Rev. B {\bf 101}, 094434 (2020).

\bibitem{SI} See Supplemental Material %at http://link.aps.org/supplemental/xxxx 
for further information about the experimental ac-susceptibility, aging, and field-cooled MH data and parameters obtained from the fitting of the experimental magnetization data using different models. 

\bibitem{MydoshTF93} J. A. Mydosh, Spin Glasses: An Experimental Introduction, Taylor \& Francis, London, (1993). 

\bibitem{MulderPRB82} C. A. M. Mulder, A. J. van Duyneveldt, and J. A. Mydosh, Frequency and field dependence of the ac susceptibility of the AuMn spin-glass, Phys. Rev. B {\bf25}, 515(R) (1982). 

\bibitem{ChakrabartyJPCM14} T. Chakrabarty, A. V. Mahajan, and S. Kundu, Cluster spin glass behavior in geometrically frustrated Zn$_3$V$_3$O$_8$, J. Phys.: Condens. Matter {\bf26}, 405601 (2014).

\bibitem{ShuklaJPCC19} R. Shukla, A. Jain, M. Miryala, K. Ueno, M. Murakami, S. M. Yusuf, and R. S. Dhaka, Spin dynamics and Unconventional Magnetism in Insulating La$_{1-2x}$Sr$_{2x}$Co$_{1-x}$Nb$_x$O$_3$, J. Phys. Chem. C {\bf123}, 222457--222469 (2019).

\bibitem{PakhiraPRB16} S. Pakhira, C. Mazumdar, R. Ranganathan, S. Giri, and M. Avdeev, Large magnetic cooling power involving the frustrated antiferromagnetic spin-glass state in R$_2$NiSi$_3$ (R = Gd, Er), Phys. Rev. B {\bf94}, 104414 (2016).

\bibitem{KhanPRB14} N. Khan, P. Mandal, and D. Prabhakaran, Memory effects and magnetic relaxation in single-crystalline La$_{0.9}$Sr$_{0.1}$CoO$_3$, Phys. Rev. B {\bf90}, 024421 (2014).

\bibitem{SouletiePRB85} J. Souletie, and J. L.Tholence, Critical slowing down in spin glasses and other glasses: Fulcher versus power law, Phys. Rev. B {\bf32}, 516--519 (1985).

\bibitem{AnandPRB12} V. K. Anand, D. T. Adroja, and A. D. Hillier, Ferromagnetic cluster spin-glass behavior in PrRhSn$_3$, Phys. Rev. B {\bf85}, 014418 (2012).

\bibitem{MalozenmoffPRB87} A. P. Malozenmoff, Random-field model of exchange anisotropy at rough ferromagnetic-antiferromagnetic interfaces, Phys. Rev. B {\bf35}, 3679 (1987).

\bibitem{MalozenmoffJAP88} A. P. Malozenmoff, Mechanism of exchange anisotropy (invited), J. Appl. Phys. {\bf63}, 3874 (1988).

\bibitem{HochstratPRB02} A. Hochstrat, Ch. Binek. and W. Kleemann, Training of exchange-bias effect in NiO-Fe heterostructures, Phys. Rev. B {\bf66}, 092409 (2002).

\bibitem{BinekPRB04} C. Binek, Training of the exchange-bias effect: A simple analytic approach, Phys. Rev. B {\bf70}, 014421(1--5) (2004).

\bibitem{MishraPRL09} S. K. Mishra, F. Radu, H. A. D$\ddot{\rm u}$rr, and W. Eberhardt, Training-Induced Positive Exchange Bias in NiFe/IrMn Bilayers, Phys. Rev. Lett. {\bf102}, 177208 (2009).

\bibitem{VenturaPRB08}   J. Ventura, J. P. Araujo, J. B. Sousa, A. Veloso, and P. P. Freitas, Training effect in specular spin valves, Phys. Rev. B {\bf77}, 184404 (2008).

\bibitem{MoutisPRB21} N. Moutis, C. Christides, I. Panagiotopoulos, and D. Niarchos, Exchange-coupling properties of La$_{1-x}$Ca$_x$MnO$_3$ ferromagnetic/antiferromagnetic multilayers, Phys. Rev. B {\bf64}, 094429 (2001).

\bibitem{HuangPRB08} X. H. Huang, J. F. Ding, G. Q. Zhang, Y. Hou, Y. P. Yao, and X. G. Li, Size-dependence exchange bias in La$_{0.25}$Ca$_{0.75}$MnO$_3$ nanoparticles, Phys. Rev. B {\bf78}, 224408 (2008).

\bibitem{PramanikPRB21} S. Gondh, M. Mishra Patidar, K. Kumar, M. P. Saravanan, V. Ganesan, and A. K. Pramanik, Large exchange bias and low-temperature glassy state in the frustrated triangular-lattice antiferromagnet Ba$_3$NiIr$_2$O$_9$, Phys. Rev. B {\bf104}, 014401 (2021).

\bibitem{KumarPRB21} R. Kumar, P. Yanda, and A. Sundaresan, Cluster-glass behavior in the two-dimensional triangular lattice Ising-spin compound Li$_2$Mn$_3$O$_7$, Phys. Rev. B {\bf103}, 214427 (2021).

\bibitem{GiriJAP12} S. K. Giri, A. Poddar, and T. K. Nath, Evidence of exchange bias effect and surface spin glass ordering in electron-doped Sm$_{0.09}$Ca$_{0.91}$MnO$_3$ nano manganites, J. Appl. Phys. {\bf112}, 113903 (2012).

\bibitem{BeanPR56} W. H. Meiklejohn and C. P. Bean, New Magnetic Anisotropy, Phys. Rev. {\bf102}, 1413 (1956).

\bibitem{NiebieskikwiatPRB05} D. Niebieskikwiat and M. B. Salamon, Intrinsic interface exchange coupling of ferromagnetic nanodomains in a charge-ordered manganite, Phys. Rev. B {\bf72}, 174422 (2005).

\bibitem{KellerPRB02} J. Keller, P. Milt\'enyi, B. Beschoten, G. G$\rm\ddot{u}$ntherodt, U. Nowak, and K. D. Usadel, Domain state model for exchange bias II. Experiments, Phys. Rev. B {\bf66}, 014431 (2002).

\bibitem{NoguesPRB96} J. Nogu\'es, C. Leighton, and I. K. Schuller, Correlation between antiferromagnetic interface coupling and positive exchange bias, Phys. Rev. B {\bf76}, 4624 (1996).

\bibitem{Kittel05} C. Kittel, Introduction to Solid State Physics, 8th ed. (Wiley, New York, 2005).

\bibitem{HeAPL09} C. He, H. Zheng, J. F. Mitchell, M. L. Foo, R. J. Cava, and C. Leighton, Low-temperature Schottky anomalies in the specific heat of LaCoO$_3$: Defect-stabilized finite spin states, Appl. Phys. Lett. {\bf94}, 102514 (2009).

\bibitem{HePRB09} C. He, S. Eisenberg, C. Jan, H. Zheng, J. F. Mitchell, and C. Leighton, Heat capacity study of magnetoelectronic phase separation in La$_{1-x}$Sr$_x$CoO$_3$ single crystals, Phys. Rev. B {\bf80}, 214411 (2009).

\bibitem{Gopal66} E. S. R. Gopal, Specific Heats at Low Temperatures (Plenum, New York, 1966).

\bibitem{GoetschPRB12} R. J. Goetsch, V. K. Anand, A. Pandey, and D. C. Johnston, Structural, thermal, magnetic, and electronic transport properties of the LaNi$_2$(Ge$_{1-x}$P$_x$)$_2$ system, Phys. Rev. B {\bf85}, 054517 (2012).

\bibitem{AjayPRB20} A. Kumar, B. Schwarz, H. Ernberg, and R. S. Dhaka, Evidence of discrete energy states and cluster-glass behavior in Sr$_{2-x}$La$_x$CoNbO$_6$, Phys. Rev. B {\bf102}, 184414 (2020).

\bibitem{GhivelderPRB99} L. Ghivelder, I. A. Castillo, M. A. Gusm\~ao, J. A. Alonso, and L. F. Cohen, Specific heat and magnetic order in LaMnO$_{3+\delta}$, Phys. Rev. B {\bf60}, 12184 (1999).

\bibitem{GhivelderJMMM98} L. Ghivelder, I. A. Castillo, N. McN. Alford, G. J. Tomka, P. C. Riedi, J. M.-Driscoll, A. K. M. A. Hossain, and L. F. Cohen, Specific heat of La$_{1-x}$Ca$_x$MnO$_{3-\delta}$, J. Magn. Magn. Mater. {\bf189}, 274 (1998).

\bibitem{WoodfieldPRL97} B. F. Woodfield, M. L. Wilson, and J. M. Byers, Low-temperature specific heat of La$_{1-x}$Ca$_x$MnO$_{3+\delta}$, Phys. Rev. Lett. {\bf78}, 3201 (1997).

\end{thebibliography}
\end{document}